\documentclass[journal]{IEEEtran}
\ifCLASSINFOpdf
\else
\fi

\usepackage{rotating}

\usepackage{url}
\usepackage{extarrows}
 \usepackage{amsmath, amssymb, latexsym}
 \usepackage{graphicx}
 \usepackage[boxed]{algorithm}
 \usepackage{color}
 \usepackage{caption}
\usepackage{algorithm}

\usepackage{amsthm}
\usepackage{multirow}

 \usepackage{psfrag}

 \newtheorem{theorem}{Theorem}
  \newenvironment{Proof}{\noindent {\bf Proof:\,}}{\hfill$\;\;\Box$ \\}
  \newtheorem{definition}{Definition}

  \newtheorem{proposition}{Proposition}
  \newtheorem{example}{Example}
  \newtheorem{lemma}{Lemma}
  \def	\Longlonghookrightarrow{\lhook\joinrel\relbar\joinrel\relbar\joinrel\relbar\joinrel\relbar\joinrel\rightarrow}
  \newcommand{\smrule}[1]{\stackrel{#1}{\Longlonghookrightarrow}}
  
  \def\by#1{\mathop{{\hbox{\setbox0=\hbox{$\scriptstyle{#1\quad}$}{$%
  					\mathrel{\mathop{\setbox1=\hbox to \wd0{\rightarrowfill}\ht1=3pt\dp1=-2pt\box1}\limits^{#1}}%
  					$}}}}}
  					
  \def\tr#1{\mathop{{\hbox{\setbox0=\hbox{$\scriptstyle{#1\quad}$}{$%
  					\mathrel{\mathop{\setbox1=\hbox to \wd0{\rightarrow}\ht1=3pt\dp1=-2pt\box1}\limits^{#1}}%
  					$}}}}}

\usepackage{url}
\urldef{\mailsa}\path| |

\begin{document}
%
\title{Reachability Analysis of Self Modifying Code}
%
%
%
\author{\IEEEauthorblockN{Tayssir Touili}
\IEEEauthorblockA{LIPN, CNRS and University Paris 13 \\
}
\and
\IEEEauthorblockN{Xin Ye}
\IEEEauthorblockA{Shanghai Key Lab. of Trustworthy Comput., ECNU and LIPN, CNRS and University Paris 13\\
}
}

\maketitle


\begin{abstract}
A Self modifying code is code that modifies its own instructions during execution time. It is nowadays widely used, especially in malware to make the code hard to analyse and to detect by anti-viruses. Thus, the analysis of such self modifying programs is a big challenge.
Pushdown systems (PDSs) is a natural model that is extensively used for the analysis of sequential programs because they allow to accurately model procedure  calls and mimic the program's stack.
In this work, we propose to extend the PushDown System model with self-modifying  rules. We call the new model Self-Modifying PushDown System
(SM-PDS). A SM-PDS is a PDS that can modify its own set of transitions during execution. We show how  SM-PDSs can be used to naturally represent self-modifying programs and provide efficient algorithms to compute the backward and forward  reachable configurations of SM-PDSs. We implemented our techniques in a tool and obtained encouraging results. In particular, we successfully applied our tool for the detection of self-modifying malware.
\end{abstract}

\section{Introduction}
Self-modifying code is code that modifies its own instructions during execution time.  It is nowadays widely used, mainly to make programs hard to understand.
For example,  self-modifying code is extensively  used to protect software intellectual property, since it makes reverse code engineering harder.
It is also abundantly used by malware writers in order to obfuscate their  malicious code   and make it  hard to analyse by static analysers and anti-viruses.

There are several kinds of implementations for self-modifying codes. 
\textbf{Packing} \cite{Debray2008} consists in applying compression techniques to make the size of the executable file smaller.
This converts  the  executable file  to a form where the executable content is hidden. 
Then, the code is "unpacked" at runtime before  execution. Such packed  code is self-modifying. 
\textbf{Encryption} is another technique to hide the code. It uses some kind of invertible operations to hide the executable code with an encryption key. 
Then, the code is "decrypted" at runtime prior to execution. Encrypted programs are self-modifying.
These  two forms of  self-modifying codes have been well studied in the litterature and could  be handled by several unpacking tools such as \cite{Selfie,unpacker}.

\begin{figure*}
	\centering
	\includegraphics[width=0.9\textwidth]{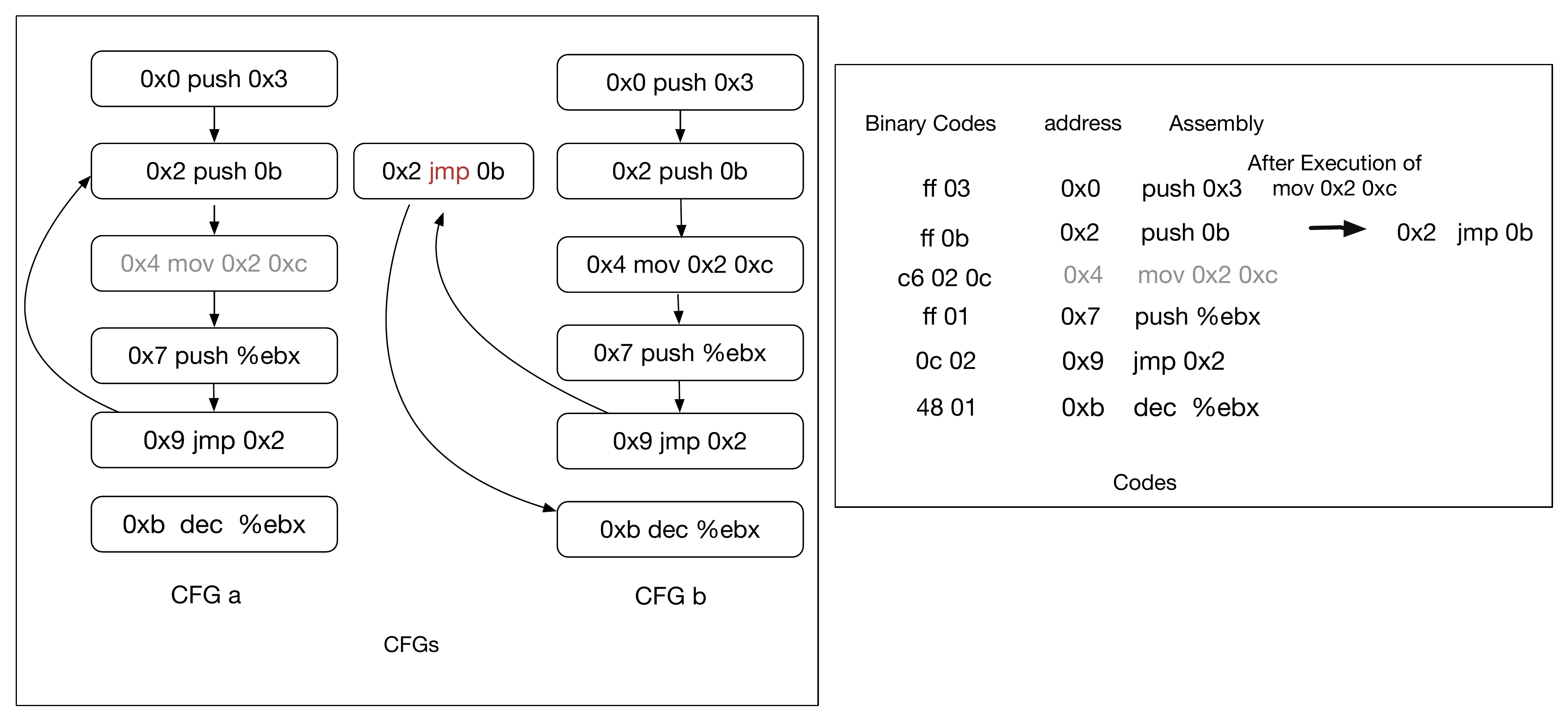}
	\caption{A Simple Example of Self-modifying Codes}\label{fig:bcc}
\end{figure*}

In this work, we consider another kind of self-modifying code,  caused by \textbf{self-modifying instructions},  where code is treated as data 
that can thus be read and written by \textbf{self-modifying instructions}.
These self-modifying instructions are usually {\bf mov} instructions, since {\bf mov} can access memory, and read and write to it. 
For example, consider the program   shown in \figurename{\ref{fig:bcc}}. For simplification matters, we suppose that  the addresses' length  is  1 byte.
The binary code is given in  the left side, while in the right side, we give its corresponding assembly code obtained by translating 
syntactically the binary code at  each address. For example, {\tt ff} is the binary code of the instruction {\tt push}, thus, the first line is translated to {\tt push 0x3}, the second line to  {\tt push 0b}, etc. 
Let us execute this code. First, we execute  {\tt push 0x3}, then {\tt push 0b}, then {\tt mov  0x2  0xc}. This last instruction will replace the first byte at address 
{\tt 0x2} by 
{\tt 0xc}. Thus, at address {\tt  0x2}, {\tt  ff 0b} is replaced by {\tt  0c 0b}. Since {\tt  0c} is the binary code of {\tt  jmp}, this means the instruction    {\tt  push 0b} is replaced by {\tt jmp  0xb}.
Therefore, this code  is self-modifying.
If we treat it   blindly, without looking at the semantics of the different instructions, we will  extract from it the Control Flow Graph {\tt CFG a},
whereas its correct  Control Flow Graph is {\tt CFG b}. You can see that the {\bf  mov} instruction was able to modify the instructions of the program successfully 
via its ability to read and write the memory.

In this paper, we consider the reachability analysis of self-modifying programs where the code is  modified by {\bf mov} instructions.
To this aim, we first need to find an adequate model for such programs.
PushDown Systems (PDS) are known to be a natural model for sequential programs \cite{Schwoon:2007vs}, as they allow to track the contexts of the different
calls in the program. Moreover,  PushDown Systems allow to  record and mimic  the  program's stack, which is very important for malware detection.
Indeed,  to check whether a program is malicious, anti-viruses start by  identifying  the calls it makes to the API functions.
To evade these checks, malware writers try to obfuscate  the  calls they make to the Operating System   by using pushes and jumps.
Thus, it is important to be able to track the stack to detect such obfuscated calls.  This is why  PushDown Systems were used in \cite{SongT12,ST13} to model binary programs 
in order to perform  malware detection.
However, these works do not consider malwares that use self-modifying code, as PushDown Systems are  not able to model  self-modifying instructions.

To overcome this limitation, we propose in this work to extend the PushDown System model with self-modifying  rules. We call the new model Self-Modifying PushDown System
(SM-PDS). 
Roughly speaking, a SM-PDS is a PDS that can modify its own set of transitions during execution.
We show how  SM-PDSs can be used to naturally represent self-modifying programs.
It turns out that SM-PDSs are equivalent to standard PDSs. We show how to translate a
 SM-PDS to a standard PDS. This translation is exponential. Thus, performing 
the reachability analysis on the equivalent PDS is not efficient. We propose then \textbf{ direct} algorithms to compute the forward ($post^*$) and backward ($pre^*$)
reachability sets for SM-PDSs. This allows to efficiently perform reachability analysis for self-modifying programs.
Our algorithms are based on (1) representing regular (potentially infinite) sets of configurations of SM-PDSs using finite state automata, and (2) applying saturation
procedures on the finite state automata in order to take into account the effect of applying the rules of the SM-PDS. 
We implemented our algorithms in a tool that takes as input either an SM-PDS or a self-modifying program.
Our experiments show that our \textbf{ direct} techniques are much more efficient than translating the SM-PDS to an equivalent PDS and then applying the standard 
reachability algorithms  for PDSs \cite{Bouajjani:1997ew,esparza,Schwoon:2007vs}.
Moreover, we  successfully applied our tool to the analysis of several self-modifying malwares.

{\bf This paper is an expanded version of  the conference paper \cite{touili2017reachability}. Compared to \cite{touili2017reachability}, in this expanded  version, we propose new  algorithms  for
  computing the forward and backward  reachable configurations for SM-PDSs, and we provide the detailed proofs
  that show the correctness of our constructions.}

\medskip
{\bf Outline.}
The rest of the paper is structured as follows: Section 2 introduces our new model and shows how to translate a SM-PDS to an equivalent pushdown system.
In Section 3, we give the  translation from  a binary code to a SM-PDS. In Section 4, we define finite automata to represent regular (potentially infinite) 
sets of configurations of  SM-PDSs.  Sections 5 and 6 give our algorithms  to compute the backward and forward reachability sets of SM-PDSs.
Section 7 describes our experiments.

\medskip
\textbf{Related Work.}

Reachability analysis of  pushdown systems  {was}  considered  in \cite{Bouajjani:1997ew,esparza}.
Our algorithms are extensions of the saturation approach of these works.

Model checking and static analysis approaches have been widely used to analyze binary programs, for instance, 
in \cite{BERGERON,Balakrishnan,SINGHf,Christodorescu2005,Kinder,ST13,Christodorescu2005}. These works cannot handle self-modifying code.

Cai et al. \cite{cai2007certified} use  a Hoare-logic-style framework to describe  self-modifying  code 
by applying local reasoning and separation logic, and   treating  program code uniformly as regular data structure.   However, \cite{cai2007certified}  requires programs to be manually annotated with invariants. 
In \cite{Debray2008}, the authors  describe  a formal semantics for  self-modifying codes, and   use that semantics to represent self-unpacking code. This work only deals with packing  and unpacking  behaviours, it   cannot capture self-modifying instructions as we do.
In \cite{bonfante2009computability}, Bonfante  et al. provide an operational semantics for self-modifying programs and  show that they  can be constructively
rewritten to a non-modifying program. 
All these specifications \cite{bonfante2009computability,cai2007certified,Debray2008} are  too abstract  to be used in practice. 

In \cite{anckaert2007model}, the authors propose a new representation of self-modifying code named State Enhanced-Control Flow Graph (SE-CFG). SE-CFG extends 
standard control flow graphs with a new data structure, keeping track of the possible states programs can reach, and with  edges that can be conditional on the state of the target memory location. 
It  is not easy to analyse a binary program only using its SE-CFG, especially that this representation does not allow to take into account the stack of the program.

\cite{Blazy2016} propose abstract interpretation techniques to compute  an over-approximation of the set of reachable states
of a self-modifying program, where for each control point of the program, an over-approximation of the memory state at this control point is provided.
\cite{roundy2010hybrid}  combine static and dynamic analysis techniques to analyse self-modifying programs.
Unlike our self-modifying pushdown systems, these techniques \cite{Blazy2016,roundy2010hybrid} cannot handle the program's stack.

Finally, unpacking binary code is considered in \cite{Coogan2009,Kang2007,Royal2006,Debray2008}. These works do not consider self-modifying 
{\bf mov} instructions.

\section{Self-modifying Pushdown Systems }\label{definition}

\subsection{Definition}
We introduce in this section our new model: Self-modifying Pushdown Systems. 

\begin{definition}
	A  Self-modifying Pushdown System (SM-PDS) is a tuple $\mathcal{P}=(P,\Gamma,\Delta,\Delta_c)$, 
	where $P$ is a finite set of control points, $\Gamma$ is a finite set of stack symbols, 
	$\Delta \subseteq (P \times \Gamma)\times (P \times \Gamma^*)$ is a finite set of transition rules, and 
	$\Delta_c \subseteq P \times (\Delta\cup\Delta_c) \times (\Delta\cup\Delta_c) \times P$ is a finite set of modifying transition rules.
	If  $((p, \gamma), (p', w)) \in \Delta$, we also write  $\langle p,\gamma \rangle \hookrightarrow \langle p',w \rangle \in \Delta$. 
	If  $(p,r_1,r_2,p') \in \Delta_c$, we also write  $p \smrule{(r_1,r_2)} p'\in \Delta_c$. A Pushdown System  (PDS) is a SM-PDS where $\Delta_c=\emptyset$.
\end{definition}

Intuitively, a Self-modifying Pushdown System is a Pushdown System   that can dynamically
modify its set of rules during the execution time: rules $\Delta$ are standard PDS transition rules, while rules $\Delta_c$
modify the current set of  transition rules:   $\langle p,\gamma \rangle \hookrightarrow \langle p',w \rangle \in \Delta$  expresses that if the 
SM-PDS is in control point $p$ and has $\gamma$
on top of its stack, then it can move to control point $p'$, pop $\gamma$ and push $w$ onto the stack, while
$p \smrule{(r_1,r_2)} p'\in \Delta_c$ expresses that when the PDS is in control point $p$, then it can move to  control point $p'$,
remove the rule $r_1$ from its current set of transition rules, and add the rule $r_2$. Formally, a configuration of a SM-PDS is a 
tuple $c=(\langle p,w\rangle,\theta)$ where $p\in P$ is the control point, $w\in\Gamma^*$ is the stack content, 
and $\theta\subseteq\Delta\cup\Delta_c$ is the current set of transition rules of the SM-PDS. $\theta$ is called the current \textbf{ phase} of the SM-PDS. When the SM-PDS is a PDS, i.e., when $\Delta_c=\emptyset$, a configuration is a tuple $c=(\langle p,w\rangle,\Delta)$, since there 
is no changing rule, so there is only one possible phase. In this case, we can also write   $c=\langle p,w\rangle$.
Let $\mathcal{C}$ be the set of configurations of a SM-PDS.
A SM-PDS defines a transition relation $\Rightarrow_{\mathcal{P}}$ between configurations as follows: Let $c=(\langle p,w\rangle,\theta)$ be a configuration,
and let $r$ be a rule in $\theta$,  then:

\begin{enumerate}
	\item if $r\in\Delta_c$ is of the form $r= p \smrule{(r_1,r_2)} p'$, such that $r_1 \in \theta$, then 
	$(\langle p, w \rangle, \theta) \Rightarrow_{\mathcal{P}} (\langle p', w \rangle,\theta')$, where 
	$\theta'=(\theta \setminus \{ r_1\}) \cup\{ r_2\}$. In other words, the transition rule $r$ updates the current set of transition rules $\theta$ by removing 
	$r_1$  from it and adding  $r_2$ to it.
	
	\item if $r\in\Delta$ is of the form $r=\langle p,\gamma \rangle \hookrightarrow \langle p',w'\rangle \in \Delta$, then 
	$(\langle p, \gamma w \rangle, \theta) \Rightarrow_{\mathcal{P}} (\langle p',w'w \rangle,\theta)$. In other words, the transition rule $r$ moves  the control point from $p$ to $p'$, 
	pops $\gamma$ from the stack and pushes $w'$ onto the stack. This transition keeps the current set of transition rules $\theta$ unchanged.
\end{enumerate}

Let  $\Rightarrow_{\mathcal{P}}^*$ be the transitive, reflexive closure of $\Rightarrow_{\mathcal{P}}$.
We define $ \stackrel{i}{\Rightarrow}$ as follows: $c \stackrel{i}{\Rightarrow} c'$ iff there exists  a sequence of configurations $c_0\Rightarrow_{\mathcal{P}} c_1 \Rightarrow_{\mathcal{P}}...\Rightarrow_{\mathcal{P}} c_i$ s.t. $c_0=c$ and $c_i=c$
Given a configuration $c$, the set of immediate predecessors (resp. successors)
of $c$ is $pre_{\mathcal{P}}(c)=\{c'\in \mathcal{C} \; : \; c' \Rightarrow_{\mathcal{P}} c\}$
(resp. $post_{\mathcal{P}}(c)=\{c'\in \mathcal{C} \; : \; c \Rightarrow_{\mathcal{P}} c'\}$). These notations
can be generalized straightforwardly to sets of configurations.
Let $pre_\mathcal{P}^*$ (resp. $post_\mathcal{P}^*$) denote the reflexive-transitive closure of $pre_\mathcal{P}$ (resp.
$post_\mathcal{P}$). We omit the subscript $\mathcal{P}$ when it is understood from the context.  

\begin{example}
	{Let $\mathcal{P}=(P,\Gamma, \Delta,\Delta_c)$ be a SM-PDS where $p=\{p_1,p_2,p_3,p_4\}$, $\Gamma=\{ \gamma_1,\gamma_2,\gamma_3\}$, $\Delta=\{r_1: \langle p_1,\gamma_1\rangle \hookrightarrow \langle p_2, \gamma_2 \gamma_1 \rangle, r_2: \langle p_2 ,\gamma_2 \rangle \hookrightarrow \langle p_3, \epsilon \rangle, r_3: \langle p_4,\gamma_1 \rangle \hookrightarrow \langle p_2, \gamma_2 \gamma_3 \rangle\}$, $\Delta_c=\{ r': p_3 \smrule{(r_1,r_3)} p_4\}$. Let $c_0=(\langle  p_1,\gamma_1 \gamma_1 \rangle, \theta_0)$ where $\theta_0=\{ r_1,r_2,r'\}$.  {Applying} rule $r_1$, we get $(\langle p_1,\gamma_1 \gamma_1 \rangle, \theta_0) \Rightarrow_{\mathcal{P}} (\langle p_2,\gamma_2\gamma_1 \gamma_1 \rangle,\theta_0)$. Then, applying rule $r_2$, we get $(\langle p_2,\gamma_2\gamma_1 \gamma_1 \rangle,\theta_0)\Rightarrow_{\mathcal{P}}  (\langle p_3,\gamma_1 \gamma_1 \rangle,\theta_0)$. Then, applying rule $r'$, we get $(\langle p_3,\gamma_1 \gamma_1 \rangle,\theta_0) \Rightarrow_{\mathcal{P}}  (\langle p_4,\gamma_1 \gamma_1 \rangle,\theta_1)$ where $r'$ is self-modifying, thus, it leads the SM-PDS from phase  $\theta_0=\{ r_1,r_2,r'\}$ to phase $\theta_1=\theta_0 \setminus \{r_1 \} \cup \{r_3\}=\{r_2,r_3,r'\}$. Then, applying rule $r_3$, we get $(\langle p_4,\gamma_1 \gamma_1 \rangle,\theta_1)
		\Rightarrow_{\mathcal{P}} (\langle p_2,\gamma_2 \gamma_3 \gamma_1 \rangle,\theta_1)$. Then, applying rule $r_2$ again, we get  $(\langle p_2,\gamma_2 \gamma_3 \gamma_1 \rangle,\theta_1) \Rightarrow_{\mathcal{P}}  (\langle p_3, \gamma_3 \gamma_1 \rangle,\theta_1)$}.
	
	\end{example}	

\subsection{From  SM-PDSs to PDSs}
\label{section-PDS}

A SM-PDS can be described by a PDS. This is due to the fact that the number of phases is finite, thus, we can encode phases in the control points of the PDS:
Let $\mathcal{P}=(P,\Gamma,\Delta,\Delta_c)$ be a SM-PDS, we compute the PDS $\mathcal{P}'=(P',\Gamma,\Delta')$ as follows:
$P'=P\times 2^{\Delta\cup\Delta_c}$. 
Initially, $\Delta'=\emptyset$. For every $\theta \in 2^{\Delta\cup\Delta_c}$,  $r\in \theta$:
\begin{enumerate}
	\item If $r =\langle p,\gamma\rangle\hookrightarrow(p',w)\in \Delta$,  we add $\langle(p,\theta),\gamma\rangle\hookrightarrow\langle(p',\theta),w\rangle$ to $\Delta'$
	\item if $r =  p \smrule{(r_1,r_2)}p'\in \Delta_c$, then  for every $\gamma \in \Gamma$, we add 
	$\langle(p,\theta),\gamma\rangle \hookrightarrow  \langle(p',\theta'),\gamma\rangle$ to $\Delta'$, where $\theta'=(\theta\setminus \{r_1\})\cup \{ r_2\}$.
\end{enumerate}

It is easy to see that:

\begin{proposition}
	$(\langle p,w\rangle,\theta)\Rightarrow_{\mathcal{P}} (\langle p',w'\rangle,\theta')$ iff
	$\langle (p,\theta),w\rangle \Rightarrow_{\mathcal{P}'}\langle (p',\theta'),w'\rangle$.
\end{proposition}
\begin{Proof}

	\noindent$\Rightarrow:$ We will show that if $(\langle p,w\rangle,\theta)\Rightarrow_{\mathcal{P}} (\langle p',w'\rangle,\theta')$, then we have $\langle (p,\theta),w\rangle \Rightarrow_{\mathcal{P}'}\langle (p',\theta'),w'\rangle$.  There are two cases depending on the form of the rule that led to this transition.
	\begin{itemize}
		
		 	\item Case $\theta=\theta':$ it means that the transition does not correspond to a self-modifying transition rule. Thus there is a rule  $r\in \theta$ of the form $r=\langle p,\gamma \rangle \hookrightarrow \langle p',u'\rangle$ that led to this transition.  Let $u$ be such that $w=\gamma u, w'=u'u$.   By the construction rule of the PDS $\mathcal{P}'$, we have $\langle (p,\theta),\gamma \rangle \hookrightarrow \langle (p',\theta),u' \rangle \in \Delta'$. Therefore, $\langle (p,\theta),\gamma u\rangle \Rightarrow_{\mathcal{P}'}\langle (p',\theta),u'u\rangle$ holds. This implies that $\langle (p,\theta),w\rangle \Rightarrow_{\mathcal{P}'}\langle (p',\theta),w'\rangle$.
		 	\item Case $\theta \neq \theta':$ it means that the transition corresponds to a self-modifying transition rule.  Thus there is a rule  $r\in \theta$ of the form $p \smrule{(r_1,r_2)} p' $ that led to this transition.  Let $u$ be such that $w=\gamma u,w'=\gamma u$.   By the construction rule of the PDS $\mathcal{P}'$, we have $\langle (p,\theta),\gamma \rangle \hookrightarrow \langle (p',\theta'),\gamma \rangle \in \Delta'$ where $\theta'=(\theta \backslash  \{r_1\}) \cup \{ r_2\}$. Therefore, $\langle (p,\theta),\gamma u\rangle \Rightarrow_{\mathcal{P}'}\langle (p',\theta'),\gamma u\rangle$ holds.  This implies that $\langle (p,\theta),w\rangle \Rightarrow_{\mathcal{P}'}\langle (p',\theta'),w'\rangle$.
		 	\end{itemize}	
	
		\noindent $\Leftarrow:$ We will show that if $\langle (p,\theta),w\rangle \Rightarrow_{\mathcal{P}'}\langle (p',\theta'),w'\rangle$, then $(\langle p,w\rangle,\theta)\Rightarrow_{\mathcal{P}} (\langle p',w'\rangle,\theta')$. Let $\gamma \in \Gamma, u ,u'\in \Gamma^*$ be such that $w=\gamma u,w'=u' u.$ There are two cases.
	\begin{itemize}
		\item Case $\theta=\theta'.$ Let $r=\langle (p,\theta),\gamma\rangle \hookrightarrow \langle (p',\theta), u'\rangle \in \Delta'$ be the rule  that led to the transition. By the construction of PDS $\mathcal{P'}$, there must exist a rule $r\in \theta$ such that $r=\langle p,\gamma \rangle \hookrightarrow \langle p',u'\rangle$. Therefore, $(\langle p,\gamma u\rangle,\theta ) \Rightarrow_{\mathcal{P}}(\langle p,u' u\rangle,\theta )$ holds. This implies that $(\langle p,w\rangle,\theta ) \Rightarrow_{\mathcal{P}}(\langle p,w'\rangle,\theta' )$.
		\item Case $\theta \neq \theta'.$  Let $r=\langle (p,\theta),\gamma\rangle \hookrightarrow \langle (p',\theta'), \gamma\rangle \in \Delta'$ be the rule leading to the transition and $u'=\gamma.$ By the construction of PDS $\mathcal{P'}$, there must exist a rule $r\in \theta$ such that $r=p\smrule{(r_1,r_2)}p'$ where $\theta'=(\theta \backslash \{r_1\}) \cup \{ r_2\}$. Therefore, $(\langle p,\gamma u\rangle,\theta ) \Rightarrow_{\mathcal{P}}(\langle p',\gamma u\rangle,\theta' )$ holds. This implies that $(\langle p,w\rangle,\theta ) \Rightarrow_{\mathcal{P}}(\langle p,w'\rangle,\theta' )$.
	\end{itemize}
\end{Proof}

Thus, we get:

\begin{theorem}
	Let $\mathcal{P}=(P,\Gamma,\Delta,\Delta_c)$ be a SM-PDS, we can compute an equivalent PDS  $\mathcal{P}'=(P',\Gamma,\Delta')$
	such that   $|\Delta'|=\big(|\Delta|+|\Delta_c|\cdot|\Gamma|\big)\cdot 2^{\mathcal{O}(|\Delta| +|\Delta_c|)}$
	and  $|P'|=|P|\cdot 2^{\mathcal{O}(|\Delta| +|\Delta_c|)}$. 
\end{theorem}
	

\subsection{From  SM-PDSs to Symbolic PDSs}
\label{section-symPDS}

Instead of recording the phases $\theta$ of the SM-PDS in the control points of the equivalent PDS,
we can have a more compact translation from SM-PDSs to \textbf{ symbolic} PDSs \cite{Schwoon:2007vs}, where each SM-PDS rule is represented by a \textbf{ single, symbolic}
transition, where the different values of the phases are encoded in a symbolic way using relations between phases:

\begin{definition}
	A symbolic pushdown system is a tuple $\mathcal{P}=(P,\Gamma,\delta)$, where $P$ is a set of control points, $\Gamma$ is the stack alphabet,
	and $\delta$ is a set of symbolic rules of the form:
	$\langle p,\gamma\rangle \smrule{R} \langle p',w\rangle$, where $R\subseteq 2^{\Delta\cup\Delta_c}\times 2^{\Delta\cup\Delta_c}$ is a relation.
\end{definition}

A symbolic PDS defines a transition relation $\leadsto_{\mathcal{P}}$ between SM-PDS configurations as follows: 
Let $c=(\langle p,\gamma w'\rangle,\theta)$ be a configuration and let $\langle p,\gamma\rangle \smrule{R} \langle p',w\rangle$ be a rule in $\delta$,  then:
$(\langle p,\gamma w'\rangle,\theta)\leadsto_{\mathcal{P}}(\langle p',ww'\rangle,\theta')$ for $(\theta,\theta')\in R$. Let $\leadsto_{\mathcal{P}}^*$
be the transitive, reflexive closure of  $\leadsto_{\mathcal{P}}$.
Then, given a SM-PDS $\mathcal{P}=(P,\Gamma,\Delta,\Delta_c)$, we can compute an equivalent symbolic PDS $\mathcal{P}'=(P,\Gamma,\Delta')$ such that:
Initially, $\Delta'=\emptyset$;
\begin{itemize}
	\item For every $\langle p,\gamma\rangle\hookrightarrow\langle p',w\rangle\in \Delta$,  add $\langle p,\gamma\rangle\smrule{R_{id}}\langle p',w\rangle$ to $\Delta'$, where $R_{id}$ is the identity relation.
	\item For every $ r= p \smrule{(r_1,r_2)}p'\in \Delta_c$ and  every $\gamma \in \Gamma$, 
	add $\langle p,\gamma\rangle\smrule{R}\langle p',\gamma\rangle $ to $ \Delta'$, where 
	$R=\{(\theta_1,\theta_2)\in 2^{\Delta\cup\Delta_c}\times 2^{\Delta\cup\Delta_c}\mid  r\in\theta_1 \text{ and } \theta_2=(\theta_1\setminus \{ r_1\})\cup \{r_2\}\}$.
\end{itemize}

It is easy to see that:

\begin{proposition}
	$(\langle p,w\rangle,\theta)\Rightarrow_{\mathcal{P}} (\langle p',w'\rangle,\theta')$ iff
	$(\langle p,w\rangle,\theta) \leadsto_{\mathcal{P'}}(\langle p',w'\rangle,\theta')$.
\end{proposition}
\begin{Proof}

\noindent
	$\Rightarrow:$ we will show that if $(\langle p,w\rangle,\theta)\Rightarrow_{\mathcal{P}} (\langle p',w'\rangle,\theta')$, then $(\langle p,w\rangle,\theta) \leadsto_{\mathcal{P'}}(\langle p',w'\rangle,\theta')$. There are two cases depending on the form of the rule that led to this transition.
	\begin{itemize}
		\item Case $\theta=\theta'$, it means that the transition does not correspond to a self-modifying transition rule. Thus there is a rule  $r\in \theta$ of the form $r=\langle p,\gamma \rangle \hookrightarrow \langle p',u'\rangle$ that led to this transition. Let $u$ be such that $w=\gamma u, w'=u'u$. By construction of the symbolic pushdown system $\mathcal{P'}$, $\langle p,\gamma \rangle \smrule{R_{id}}\langle p',u'\rangle \in \Delta'$, therefore, $(\langle p,\gamma u\rangle,\theta) \leadsto_{\mathcal{P'}}(\langle p',u' u\rangle,\theta)$ holds. This implies that $(\langle p,w\rangle,\theta) \leadsto_{\mathcal{P'}}(\langle p',w'\rangle,\theta').$

		\item  Case $\theta\neq\theta'$, it means that the transition corresponds to a self-modifying transition rule. Thus there is a rule  $r\in \theta$ of the form $r= p \smrule{(r_1,r_2)} p'$ that led to this transition and $\theta'=(\theta \backslash \{r_1\}) \cup \{ r_2\}$. Let $u$ be such that $w=\gamma u, w'=\gamma u$. By construction of the symbolic pushdown system $\mathcal{P}'$, $\langle p,\gamma \rangle \smrule{R}\langle p',\gamma\rangle \in \Delta'$ and $R=\{(\theta,\theta')\in 2^{\Delta\cup\Delta_c}\times 2^{\Delta\cup\Delta_c}\mid  r\in\theta \text{ and } \theta'=(\theta\setminus \{ r_1\})\cup \{r_2\}\}$, therefore, $(\langle p,\gamma u\rangle,\theta) \leadsto_{\mathcal{P'}}(\langle p',\gamma u\rangle,\theta')$ holds.  This implies that $(\langle p,w\rangle,\theta) \leadsto_{\mathcal{P'}}(\langle p',w'\rangle,\theta').$	
		\end{itemize}
		\noindent
	$\Leftarrow:$ we will show that  if $(\langle p,w\rangle,\theta) \leadsto_{\mathcal{P'}}(\langle p',w'\rangle,\theta')$, then $(\langle p,w\rangle,\theta)\Rightarrow_{\mathcal{P}} (\langle p',w'\rangle,\theta')$. Let $\gamma \in \Gamma, u,u'\in \Gamma^*$ be such that $w=\gamma u, w'=u' u$. There are two cases.
	\begin{itemize}
		\item Case $\theta=\theta'$. Let $\langle p,\gamma \rangle \smrule{R_{id}} \langle p', u'\rangle \in \Delta'$ be the rule applied to this transition. By the construction of the symbolic pushdown system $\mathcal{P}'$, there must exist a rule $r\in \theta$ s.t. $r=\langle p,\gamma \rangle \hookrightarrow\langle p',u'\rangle \in \Delta$. Therefore, $(\langle p,\gamma u\rangle,\theta)\Rightarrow_{\mathcal{P}} (\langle p', u'u\rangle,\theta)$ holds.  This implies that $(\langle p,w\rangle,\theta) \Rightarrow_{\mathcal{P}}(\langle p',w' \rangle,\theta')$.
		\item Case  $\theta \neq \theta'$. Let $\langle p,\gamma \rangle \smrule{R} \langle p', \gamma\rangle \in \Delta'$ be the rule applied to this transition with $w'=\gamma u$. By the construction of  {the} symbolic pushdown system $\mathcal{P}'$, there must exist a rule $r\in \theta$ of the form $r=p\smrule{(r_1,r_2)} p'\in \Delta_c$ s.t. $R=\{(\theta_1,\theta_2)\in 2^{\Delta\cup\Delta_c}\times 2^{\Delta\cup\Delta_c}\mid  r\in\theta_1 \text{ and } \theta_2=(\theta_1\setminus \{ r_1\})\cup \{r_2\}\}$. Therefore, $\theta'=(\theta\backslash \{ r_1\}) \cup \{ r_2\}$ and $(\langle p,\gamma u\rangle,\theta) \leadsto_{\mathcal{P'}}(\langle p',\gamma u\rangle,\theta')$ hold.  This implies that $(\langle p,w\rangle,\theta) \leadsto_{\mathcal{P'}}(\langle p',w'\rangle,\theta').$	
	\end{itemize}
\end{Proof}

Thus, we get:

\begin{theorem}
	Let $\mathcal{P}=(P,\Gamma,\Delta,\Delta_c)$ be a SM-PDS, we can compute an equivalent symbolic  PDS  $\mathcal{P}'=(P',\Gamma,\Delta')$
	such that $|P'|=|P|$, $|\Delta'|=|\Delta|+|\Delta_c|\cdot |\Gamma|$, and the size of the relations used in the symbolic transitions is
	$2^{\mathcal{O}(|\Delta|+|\Delta_c|)}$.
\end{theorem}

\section{Modeling self-modifying code with SM-PDSs}

\newcommand{\States}{\textbf{States}}
\newcommand{\Reg}{\textbf{R}}
\newcommand{\Val}{\textbf{Val}}
\newcommand{\Mem}{\textbf{Mem}}
\newcommand{\EXP}{\textbf{EXP}}
\newcommand{\Z}{\mathbb{Z}}
\newcommand{\Orac}{\mathcal{O}}

\subsection{Self-modifying instructions}
\label{section-instructions}
There are different techniques to implement self-modifying code. 
We consider in this work  code that uses self-modifying instructions.
These are instructions that can access the memory locations  and write onto them, thus changing the instructions that    are in these memory locations.
In assembly, the only instructions that can do this are the {\bf mov} instructions.   In this case, the self-modifying instructions are of the form {\tt mov} $l~v$, where $l$ is a location of the program that stores executable data and $v$ is a value. This instruction
replaces the value at   location $l$ (in the binary code)  with the  value $v$.
This means if at location $l$ there is a binary value $v'$ that is involved in  an assembly instruction $i_1$, and if by replacing $v'$ by $v$, we 
obtain a new assembly instruction $i_2$, then 
the instruction $i_1$ is replaced by $i_2$.   E.g., {\tt ff} is the binary code of {\tt push}, {\tt 40} is the binary code of {\tt inc},
{\tt 0c} is the binary code of {\tt jmp}, {\tt c6} is the binary code of {\tt mov},  etc.
Thus, if we have  {\tt mov} $ l$ {\tt ff}, and if at location $l$ there was initially the value 40 01 (which corresponds to the assembly instruction
inc \%edx), then 40 is replaced by   {\tt ff}, which means the instruction  {\tt inc \%edx} is replaced by {\tt push 01}.
If at location $l$ there was initially the value c6 01 02  (which corresponds to the assembly instruction
{\tt mov edx 0x2}), then c6 is replaced by   {\tt ff}, which means the instruction  {\tt mov edx 0x2} is replaced by {\tt  push 02}.

Note that if the instructions $i_1$ and $i_2$ do not have the same number of operands, then  {\tt mov} $l~v$ will, in addition to  replacing    $i_1$ by  $i_2$,
change several other instructions that follow    $i_1$. Currently, we cannot handle this case, thus we   assume that $i_1$ and $i_2$ have the same number of operands.

Note also that  $mov~l~v$   is self-modifying only if $l$ is a location of the program that stores executable data, otherwise, it is not;
e.g., $mov~eax~v$ does not change the instructions of the program, it just writes the value $v$ to the register $eax$.
Thus, from now on,  by self-modifying instruction, we mean an instruction of the form  $mov~l~v$, where $l$ is a location 
of the program that stores executable data. 
Moreover, to ensure that only one instruction is modified, 
		we assume that the corresponding instructions $i_1$ and   $i_2$ have the same number of operands.

\subsection{From self-modifying code to  SM-PDS}
\label{prog-pds}

We show in what follows how to build a SM-PDS from a binary program.
We suppose we are given  an oracle  $\mathcal{O}$  that  extracts from the binary code  a corresponding  assembly  program, together with   informations about the values of
the registers and the  memory locations at each control point   of the program.
In our implementation, we use Jakstab \cite{jakstab} to get this oracle.
We translate the assembly program  into a  self-modifying pushdown system
where the control locations store the control points of the binary program and
the stack  memics the program's stack. The non self-modifying instructions  of the program define the rules $\Delta$ of the SM-PDS (which are standard PDS rules),
and can be obtained following the translation of \cite{SongT12} that models non self-modifying instructions  of the program by a PDS.

As for the  self-modifying instructions of the program, they define 
the set of  changing rules $\Delta_c$. As explained above, these are  
instructions  of the form $mov~l~v$, where $l$ is a location of the program that stores executable data. This instruction
replaces the value  at   location $l$ (in the binary code)  with the  value $v$.  
Let  $i_1$ be the  initial instruction involving the  location $l$,  and let  $i_2$ 
be the new instruction  involving the  location $l$,  after applying the $mov~l~v$ instruction. 
As mentioned  previously, we assume that  $i_1$ and   $i_2$ have the same number of operands (to ensure that only one instruction is modified).
Let $r_1$ (resp. $r_2$) be  the SM-PDS rule corresponding to
the instruction  $i_1$  (resp.  $i_2$).
Suppose from control point $n$ to $n'$, we have this $mov~l~v$ instruction, 
then we add $n \smrule{(r_1, r_2)}n'$ to $\Delta_c$. This is the SM-PDS rule corresponding to the instruction $mov~l~v$ at control point $n$.

\section{Representing infinite sets of configurations of a SM-PDS}

Multi-automata were introduced in \cite{Bouajjani:1997ew,esparza} to finitely represent regular infinite sets of configurations of a PDS.
A configuration $c=(\langle p,w\rangle,\theta)$ of a SM-PDS  involves a PDS configuration $\langle p,w\rangle$,
together with the current set of  transition rules (phase) $\theta$. To finitely represent regular infinite sets of such configurations, we 
extend multi-automata in order to take into account the phases $\theta$:

\begin{definition}
 \label{labelledauto}
 	 Let $\mathcal{P}=(P,\Gamma,\Delta,\Delta_c)$ be a SM-PDS. A  $\mathcal{P}$-automaton is a tuple  $\mathcal{A}=(Q,\Gamma,T,P,F)$ where  $\Gamma$ is the automaton alphabet, $Q$ is a finite set of states,
 $P\times 2^{\Delta \cup \Delta_c}\subseteq Q$ is the set of initial states,   $T \subset Q \times \big((\Gamma\cup\{\epsilon\}) \big)\times Q $ is the set of transitions and $F\subseteq Q$ is the set of final states.
\end{definition}

\noindent
If $\big(q,\gamma,q'\big)\in T$, we write $q \by{\gamma}_T q'$. 	We extend this notation in the obvious manner to sequences of symbols:
	(1) $\forall q \in Q, q \by{\epsilon}_T q$, and 
	(2) $\forall q, q' \in Q, \theta'\in 2^{\Delta\cup \Delta_c},  \forall \gamma \in \Gamma\cup\{\epsilon\},  
	\forall w \in \Gamma^* \text{ for }w=\gamma_0\gamma_1\cdots\gamma_n,  q\by{\gamma w}_T q'$ iff $\exists q'' \in Q, q \by{\gamma}_T q'' \; \mbox{and} \;q'' \by{w}_T q'$.  If $q\xrightarrow{w}_{T}q'$ holds, we say that $q\xrightarrow{w}_{T}q'$ is a path of $\mathcal{A}$.  
	A configuration $(\langle p, w \rangle, \theta)$ is  accepted by  $\mathcal{A}$ iff  $\mathcal{A}$ contains a path
	$(p,\theta){\by{\gamma_0}}_T q_1{\by{\gamma_1}}_T q_2\cdots q_n{\by{\gamma_n}}_T q$ where $q\in F$.	Let $L(\mathcal{A})$ be the set of configurations accepted by $\mathcal{A}$.
	Let $\mathcal{C}$ be a  set of configurations of the  SM-PDS $\mathcal{P}$. 
	$\mathcal{C}$ is regular  if there exists a $\mathcal{P}$-automaton $\mathcal{A}$ such that  $\mathcal{C}=L(\mathcal{A})$

\section{Efficient computation of $pre^*$ images}
Let $\mathcal{P}=(P,\Gamma,\Delta,\Delta_c)$ be a SM-PDS, and let $\mathcal{A}=(Q,\Gamma,T,P,F)$ be a  $\mathcal{P}$-automaton that represents a regular
set of configurations $\mathcal{C}$ ( $\mathcal{C}=L(\mathcal{A})$). To compute $pre^*( \mathcal{C})$, one can use the translation of Section \ref{section-PDS} to compute an equivalent PDS, and then apply the algorithms of  \cite{Bouajjani:1997ew,esparza}. This procedure is too complex since the size of the obtained PDS is huge. One can also use the translation of Section \ref{section-symPDS} to compute an equivalent symbolic PDS, and then use the algorihms of  \cite{Schwoon:2007vs}.
However, this procedure is not optimal neither since the number of elements of the relations considered in the rules of the symbolic  PDSs are huge.
We present in this section a \textbf{ direct} and   \textbf{ more efficient} algorithm that computes $pre^*( \mathcal{C})$ without any need to translate the SM-PDS to an equivalent PDS or symbolic PDS.
We assume w.l.o.g. that $\mathcal{A}$ has no transitions leading to an initial state.
We also assume that the self-modifying rules $r=p \smrule{(r_1,r_2)} p'$ in $\Delta_c$ are such that $r\neq r_1$. This is not a restriction since 
a rule of the form $r=p \smrule{(r,r_2)} p'$ can be replaced by these  rules that meet this constraint: 
$r= p \smrule{(r_{\bot},r_{\bot})} p_i$ and $p_i \smrule{(r,r_2)} p'$, where $r_{\bot}$ is a new fake rule that we can add to all phases.

The construction of $\mathcal{A}_{pre^*}$ follows the same idea  as  
for standard  pushdown systems (see \cite{Bouajjani:1997ew,esparza}).
It consists in adding iteratively new transitions to the automaton $\mathcal{A}$ according
to \textbf{ saturation} rules (reflecting the backward application of the transition
rules in the system), while the set of states remains unchanged.  Therefore, let $\mathcal{A}_{pre^*}$  be the $\mathcal{P}$-automaton 
$(Q,\Gamma,T',P,F)$, where  $T'$ is computed using the following saturation rules: 
initially $T'=T$.

\begin{enumerate}
\item[$\alpha_1$:] If $r=\langle p,\gamma \rangle \hookrightarrow \langle p_1,w \rangle \in \Delta$, where $w\in \Gamma^*$. For every $\theta \subseteq \Delta \cup \Delta_c$ s.t. $r\in \theta$, 
if  there exists in $T'$ a path $\pi=(p_1,\theta)\by{w}_{T}q$, then add $((p,\theta),\gamma,q)$ to $T'$.

\item[$\alpha_2$:]  if $r=p\smrule{(r_1,r_2)} p_1 \in \Delta_c$ for every $\theta \subseteq \Delta \cup \Delta_c$ s.t. $r\in \theta, r_2 \in\theta$  and for every $\gamma \in \Gamma$, if there exists in $T'$ a transition   $t=(p_1,\theta)\by{\gamma}_{T}q$,  then add $((p,\theta'),\gamma,q)$ to $T'$ where $ \theta=(\theta'\setminus \{r_1\}) \cup \{r_2\}$.
\end{enumerate}

\noindent
The  procedure above  terminates since there is a finite number of states and phases.

\noindent
Let us explain intuitively the role of the saturation rule ($\alpha_1$). Let $r=\langle p,\gamma \rangle \hookrightarrow \langle p',w \rangle \in \Delta$.
Consider a path in the automaton of the form $(p',\theta')\by{w}_{T'}q\by{w'}_{T'}q_F$, where $q_F\in F$. This means, by definition of  $\mathcal{P}$-automata,  that the configuration $c=(\langle p',ww'\rangle,\theta')$ is accepted by $\mathcal{A}_{pre^*}$.
If $r$ is in $\theta'$, then the configuration $c'=(\langle p,\gamma w'\rangle,\theta')$ is a predecessor of $c$. Therefore, it should be added to 
$\mathcal{A}_{pre^*}$. This configuration is accepted by the run $(p,\theta')\by{\gamma}_{T'}q\by{w'}_{T'}q_F$ added by rules  ($\alpha_1$).

Rule ($\alpha_2$) deals with modifying rules: Let  $r=p\smrule{(r_1,r_2)} p' \in \Delta_c$. 
Consider a path in the automaton of the form $(p',\theta')\by{\gamma}_{T'}q\by{w'}_{T'}q_F$, where $q_F\in F$. 
This means, by definition of $\mathcal{P}$-automata,  that the configuration $c=(\langle p',\gamma w'\rangle,\theta')$ is accepted by $\mathcal{A}_{pre^*}$.
If $r$ and $r_2$ are  in $\theta'$, then the configuration $c'=(\langle p,\gamma w'\rangle,\theta)$ is a predecessor of $c$, where  
$\theta'=(\theta\setminus \{r_1\}) \cup \{r_2\}$. Therefore, it should be added to 
$\mathcal{A}_{pre^*}$. This configuration is accepted by the run $\pi'=(p,\theta)\by{\gamma}_{T'}q\by{w'}_{T'}q_F$ added by rules  ($\alpha_2$).

\begin{figure}
\centering
\includegraphics[width=8.5cm,height=4.5cm]{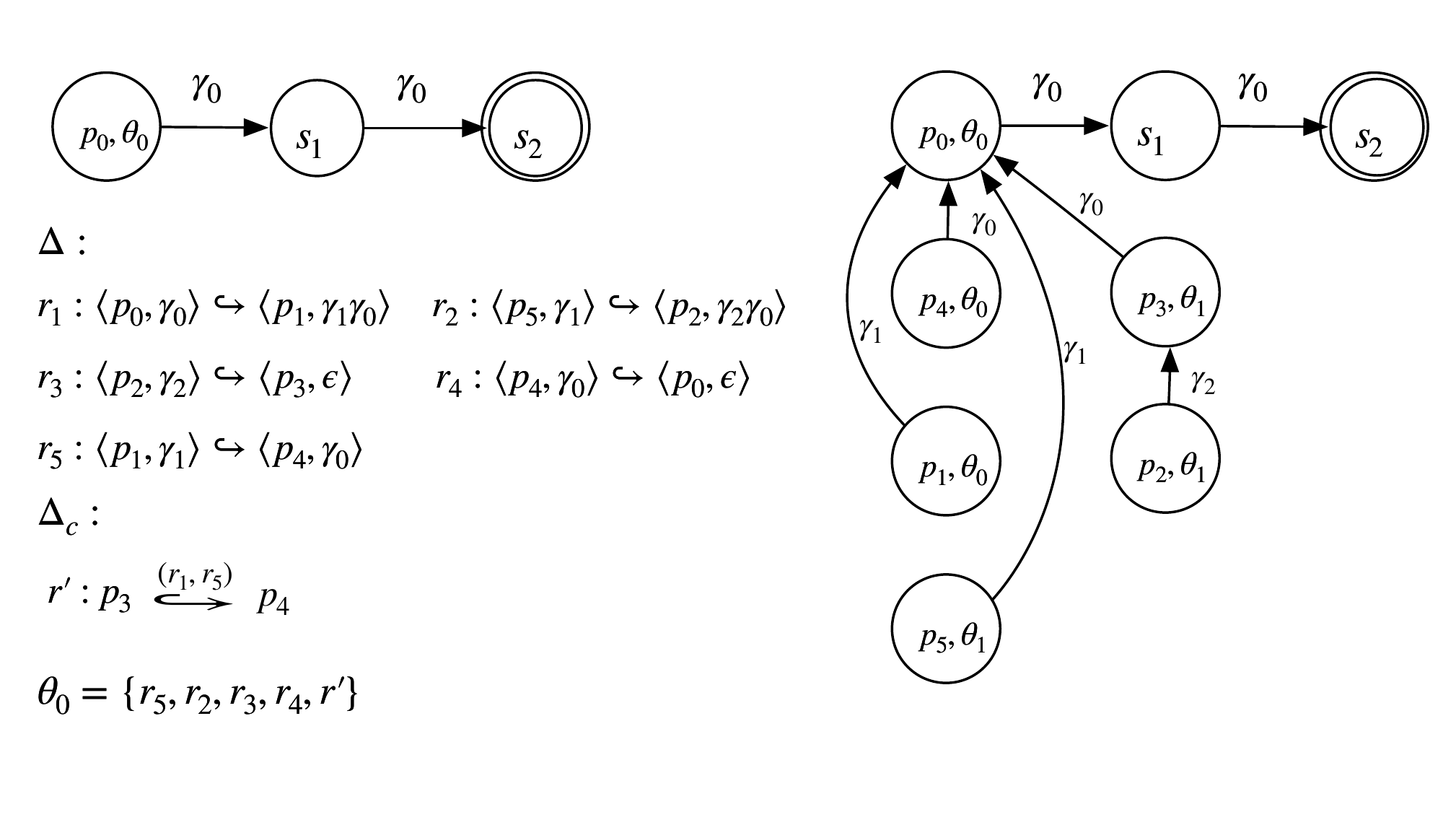}
\caption{The automata $\mathcal{A}$ (left) and $\mathcal{A}_{pre^*}$ (right)}\label{fig:preex}
\end{figure}
\medskip
Thus, we can show that:
\begin{theorem}
\label{pre}
	$\mathcal{A}_{pre^*}$ recognizes  $pre^*({L(A)})$.
\end{theorem}

{Before proving this theorem, let us illustrate the construction on 2 examples.}

\begin{example}
	{Let us illustrate the procedure by an example. Consider the SM-PDS with control points $P=\{ p_0,p_1, p_2,p_3,p_4,p_5 \}$ and $\Delta, \Delta_c$ as shown in the left half of Fig. \ref{fig:preex}.  Let $\mathcal{A}$ be the automaton that accepts the set $C=\{(\langle p_0,\gamma_0 \gamma_0 \rangle,\theta_0)\},$ also shown on the left where $(p_0,\theta_0)$ is the initial state and $s_2$ is the final state. The result of the algorithm is shown in the right half of Fig. \ref{fig:preex}. The result is obtained through the following steps:} 
	\begin{enumerate}
		\item  { First, we note that $(p_0,\theta_0) \xrightarrow{\epsilon}_{T'}(p_0,\theta_0)$ holds. Since $\langle p_0,\epsilon \rangle$ occurs on the right hand side of rule $r_4$ and $r_4 \in \theta_0$,  then Rule $(\alpha_1)$ adds the transition $(p_4,\theta_0)\xrightarrow{\gamma_0} (p_0,\theta_0)$ to $T'$.}
		\item  { Now that we have $(p_4,\theta_0)\xrightarrow{\gamma_0}_{T'} (p_0,\theta_0)$, since $r_5\in \theta_0$, Rule $(\alpha_1)$ adds  $(p_1,\theta_0)\xrightarrow{\gamma_1} (p_0,\theta_0)$ to $T'$.}
		\item  { Since we have $(p_4,\theta_0)\xrightarrow{\gamma_0}_{T'} (p_0,\theta_0)$, the self-modifying transition $r'\in \theta_0$ can be applied. Thus, Rule $(\alpha_2)$ adds $(p_3,\theta_1) \xrightarrow{\gamma_0}(p_0,\theta_0)$ to $T'$ where $\theta_1 =(\theta_0 \setminus \{r_5\} )\cup \{r_1\} =\{r_1,r_2,r_3,r_4,r'\}$.}
		\item  { Since $(p_3,\theta_1) \xrightarrow{\epsilon}(p_3,\theta_1)$ and $r_3 \in \theta_1$, Rule $(\alpha_1)$ adds $(p_2,\theta_1)\xrightarrow{\gamma_2} (p_3,\theta_1)$ to $T'$.}
		\item  { Then, there is a path  $(p_2,\theta_1)\xrightarrow{\gamma_2}_{T'} (p_3,\theta_1) \xrightarrow{\gamma_0}_{T'}(p_0,\theta_0)$. Since $\langle p_2,\gamma_2 \gamma_0 \rangle$ occurs on the right hand side of $r_2$ and $r_2 \in \theta_1$, then Rule $(\alpha_1)$ adds the transition $(p_5,\theta_1) \xrightarrow{\gamma_1}(p_0,\theta_0)$ to $T'$.}
		\item  { No further additions are possible. Thus, the procedure terminates.}
 	\end{enumerate}
\end{example}
\begin{figure*}[htbp]
\centering
\includegraphics[width=13cm,height=5cm]{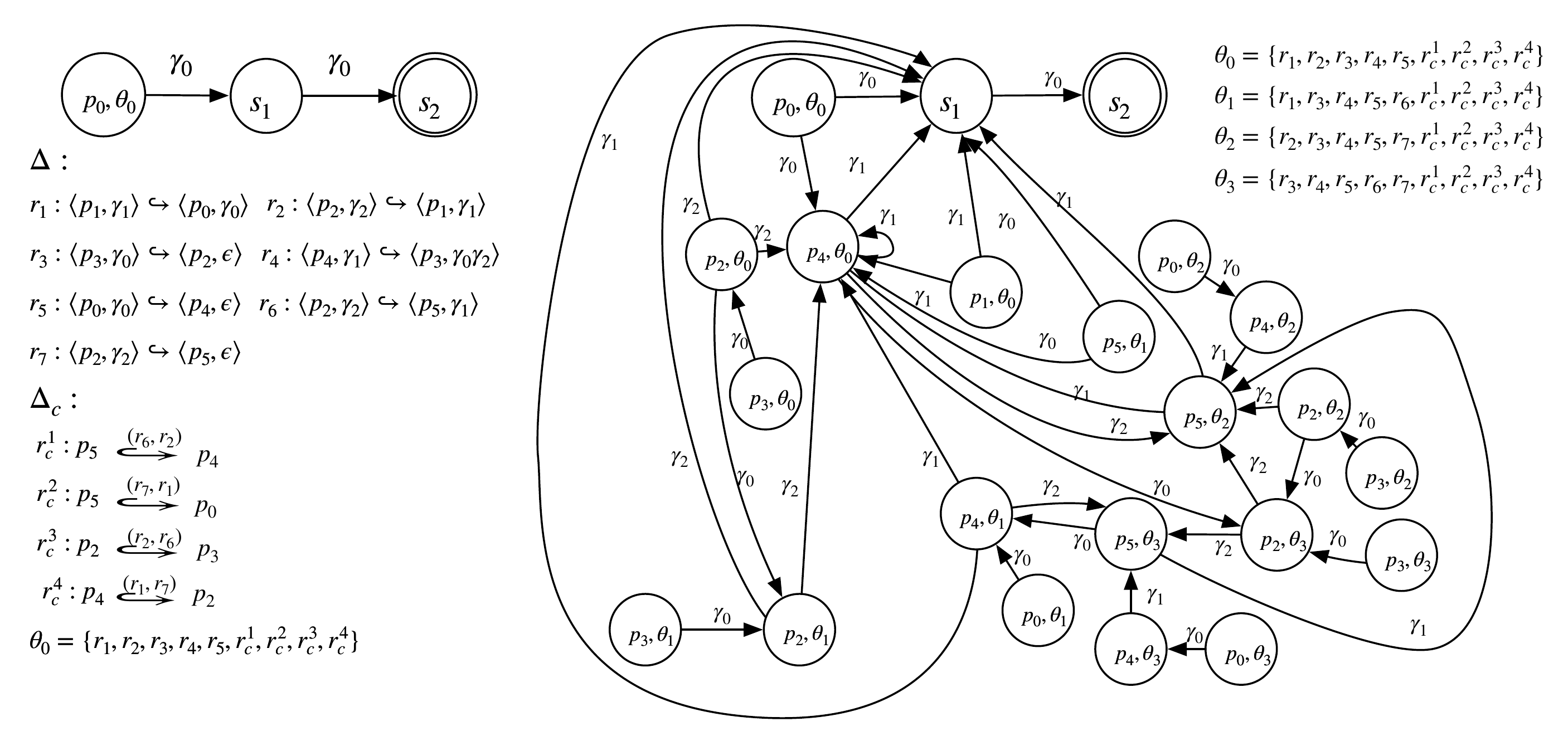}
\caption{The automata $\mathcal{A}$ (left) and $\mathcal{A}_{pre^*}$ (right)}\label{fig:compex}
\end{figure*}
\begin{example}
{
	Let us give another example. Consider the SM-PDS with control points $P=\{ p_1,p_2,p_3,p_4,p_5\}$ and $\Delta,\Delta_c$ as shown in the left half of Fig. \ref{fig:compex}. Let $\mathcal{A}$ be the automaton that accepts the set $C=\{(\langle p_0,\gamma_0 \gamma_0\rangle,\theta_0)\}$ where $(p_0,\theta_0)$ is the initial state and $s_2$ is the  {final} state as shown on the left. The result $\mathcal{A}_{pre^*}$ of the algorithm is on the right half of Fig. \ref{fig:compex}. The result is obtained through the following steps:
	\begin{enumerate}
		\item Since $(p_0,\theta_0) \xrightarrow{\gamma}_{T'}s_1$ and  $r_1\in \theta_0$, then Rule $(\alpha_1)$ adds $(p_1,\theta_0)\xrightarrow{\gamma_1} s_1$ to $T'$.
		\item Since $(p_1,\theta_0)\xrightarrow{\gamma_1}_{T'} s_1$  and $r_2\in \theta_0$, Rule $(\alpha_1)$ adds the transition $(p_2,\theta_0)\xrightarrow{\gamma_2} s_1$ to $T'$.
		\item Since $(p_2,\theta_0)\xrightarrow{\epsilon}_{T'} (p_2,\theta_0)$ and $r_3\in \theta_0$, Rule $(\alpha_1)$ adds the transition $(p_3,\theta_0)\xrightarrow{\gamma_0} (p_2,\theta_0)$ to $T'$.
		\item Then, there is a path $(p_3,\theta_0)\xrightarrow{\gamma_0}_{T'} (p_2,\theta_0) \xrightarrow{\gamma_2}_{T'}s_1$  and  $r_4\in \theta_0$, Rule $\alpha_1$ adds the transition $(p_4,\theta_0)\xrightarrow{\gamma_1}s_1$ to $T'$.
		\item Because $(p_4,\theta_0)\xrightarrow{\epsilon}_{T'} (p_4,\theta_0)$ and $r_5 \in \theta_0$, Rule $(\alpha_1)$ adds the transition $(p_0,\theta_0) \xrightarrow{\gamma_0}(p_4,\theta_0)$ to $T'$.
		\item Since $(p_0,\theta_0) \xrightarrow{\gamma_0}_{T'}(p_4,\theta_0)$ and $r_1\in \theta_0$, Rule $(\alpha_1)$  adds the transition $(p_1,\theta_0)\xrightarrow{\gamma_1}(p_4,\theta_0)$ to $T'$. Then, since $r_2\in \theta_0$, Rule $(\alpha_1)$ adds the transition $(p_2,\theta_0)\xrightarrow{\gamma_2}(p_4,\theta_0)$ to $T'$.
		\item Since there is  a path $(p_3,\theta_0)\xrightarrow{\gamma_0}_{T'}(p_2,\theta_0)\xrightarrow{\gamma_2}_{T'}(p_4,\theta_0)$ and $r_4\in \theta_0$, Rule $(\alpha_1)$ adds $(p_4,\theta_0)\xrightarrow{\gamma_1}(p_4,\theta_0)$ to $T'$.
		\item Since $(p_4,\theta_0)\xrightarrow{\gamma_1}_{T'}s_1,(p_4,\theta_0)\xrightarrow{\gamma_1}_{T'}(p_4,\theta_0)$ and $r_c^1,r_2\in \theta_0$, Rule $(\alpha_2)$ adds $(p_5,\theta_1)\xrightarrow{\gamma_1}s_1$ and $(p_5,\theta_1)\xrightarrow{\gamma_1}(p_4,\theta_0)$ to $T'$ where $\theta_1=(\theta_0 \setminus \{ r_2\}) \cup r_6=\{r_1,r_3,r_4,r_5,r_6,r_c^1,r_c^2,r_c^3,r_c^4\}$. For the same reason, since  $(p_0,\theta_0) \xrightarrow{\gamma_0}_{T'}(p_4,\theta_0)$, $(p_0,\theta_0) \xrightarrow{\gamma}_{T'}s_1$ and $r_1,\in \theta_1$, $r_c^2\in \theta_0$, Rule $(\alpha_2)$  adds the transitions $(p_5,\theta_2)\xrightarrow{\gamma_0}(p_4,\theta_0)$ and $(p_5,\theta_2)\xrightarrow{\gamma_0} s_1$ to $T'$ where $\theta_2=(\theta_0\setminus \{r_1\}) \cup \{r_7\}=\{r_2,r_3,r_4,r_5,r_7,r_c^1,r_c^2,r_c^3,r_c^4\}.$
		\item Since $(p_5,\theta_1)\xrightarrow{\gamma_1}_{T'}s_1$, $(p_5,\theta_1)\xrightarrow{\gamma_1}_{T'}(p_4,\theta_0)$ and $r_6\in \theta_1$, Rule $(\alpha_1)$ adds the transitions $(p_2,\theta_1)\xrightarrow{\gamma_2} s_1$ and $(p_2,\theta_1)\xrightarrow{\gamma_2} (p_4,\theta_0)$ to $T'$.
		\item Since $(p_2,\theta_1)\xrightarrow{\epsilon}_{T'}(p_2,\theta_1)$  and $r_3 \in \theta_1$, Rule $(\alpha_1)$ adds $(p_3,\theta_1)\xrightarrow{\gamma_0}(p_2,\theta_1)$.
		\item Because there are paths $(p_3,\theta_1)\xrightarrow{\gamma_0}_{T'}(p_2,\theta_1)\xrightarrow{\gamma_2}_{T'}(p_4,\theta_0)$ and $(p_3,\theta_1)\xrightarrow{\gamma_0}_{T'}(p_2,\theta_1)\xrightarrow{\gamma_2}_{T'}s_1$,  Rule $(\alpha_1)$
   adds the transitions $(p_4,\theta_1)\xrightarrow{\gamma_1}(p_4,\theta_0)$ and $(p_4,\theta_1)\xrightarrow{\gamma_1}s_1$ to $T'$.
		\item Since $(p_4,\theta_0)\xrightarrow{\epsilon}_{T'} (p_4,\theta_0)$ and $r_5 \in \theta_1$, Rule $(\alpha_1)$ adds $(p_0,\theta_1)\xrightarrow{\gamma_0}(p_4,\theta_1)$.
		\item Now we have $(p_0,\theta_1)\xrightarrow{\gamma_0}_{T'}(p_4,\theta_1)$ and $r_c^2,r_1\in \theta_1$, Rule $(\alpha_2)$ adds the transition $(p_5,\theta_3)\xrightarrow{\gamma_0}(p_4,\theta_1)$  to $T'$ where $\theta_3=\{r_3,r_4,r_5,r_6,r_7,r_c^1,r_c^2,r_c^3,r_c^4\}$. For the same reason, since $(p_3,\theta_1)\xrightarrow{\gamma_0}(p_2,\theta_1)$ and $r_c^3,r_6\in \theta_1$, Rule $\alpha_2$ adds the transition $(p_2,\theta_0)\xrightarrow{\gamma_0}(p_2,\theta_1)$ to $T'$ because $\theta_0=(\theta_1 \setminus \{r_6\})\cup \{r_2\}$.
		\item Since $(p_5,\theta_3)\xrightarrow{\epsilon}_{T'}(p_5,\theta_3)$ and $r_7\in \theta_3$, Rule $(\alpha_1)$  adds the transition $(p_2,\theta_3)\xrightarrow{\gamma_2}(p_5,\theta_3)$ to $T'$.
		\item Because $(p_2,\theta_3)\xrightarrow{\epsilon}_{T'}(p_2,\theta_3)$  and $r_3\in \theta_3$, Rule ($\alpha_1$) adds the transition  $(p_3,\theta_3)\xrightarrow{\gamma_0}(p_2,\theta_3)$ to $T'$. Then, since there is  a  path $(p_3,\theta_3)\xrightarrow{\gamma_0}_{T'}(p_2,\theta_3)\xrightarrow{\gamma_2}_{T'}(p_5,\theta_3)$ and $r_4 \in \theta_3$, Rule $(\alpha_1)$ adds the transition $(p_4,\theta_3)\xrightarrow{\gamma_1}(p_5,\theta_3)$ to $T'$. Then, since $r_5\in \theta_3$, Rule $(\alpha_1)$ adds the transition $(p_0,\theta_3)\xrightarrow{\gamma_0}(p_4,\theta_3)$ to $T'$.
		\item  Since $(p_3,\theta_3)\xrightarrow{\gamma_0}_{T'}(p_2,\theta_3)$ and $r_c^3 \in \theta_3$, Rule $(\alpha_2)$  adds  the transition $(p_2,\theta_2)\xrightarrow{\gamma_0}_{T'}(p_2,\theta_3)$ to $T'$ where $(\theta_3\setminus\{r_6\})\cup \{r_2\}=\theta_2$. Meanwhile,  since  $(p_2,\theta_3)\xrightarrow{\gamma_2}(p_5,\theta_3)$  and $r_c^4, r_7 \in \theta_3$, Rule $(\alpha_2)$ adds the transition $(p_4,\theta_1)\xrightarrow{\gamma_2}(p_5,\theta_3)$ to $T'$ where $(\theta_3\setminus \{ r_7\})\cup \{r_1\}=\theta_1$.
		\item Because $r_7\in \theta_2$ and $(p_5,\theta_2)\xrightarrow{\epsilon}_{T'}(p_5,\theta_2)$, Rule ($\alpha_1$) adds the transition  $(p_2,\theta_2)\xrightarrow{\gamma_2}(p_5,\theta_2)$ to $T'$.
		\item  Since $(p_2,\theta_2)\xrightarrow{\epsilon}_{T'}(p_2,\theta_2)$ and $r_3 \in \theta_2$, Rule $(\alpha_1)$ adds $(p_3,\theta_2)\xrightarrow{\gamma_0}_{T'}(p_2,\theta_2)$ to $T'$. Then, there is a path  $(p_3,\theta_2)\xrightarrow{\gamma_0\gamma_2}_{T'}^* (p_5,\theta_2)$, since $r_4\in \theta_2$,  Rule $(\alpha_1)$ adds the transition $(p_4,\theta_2)\xrightarrow{\gamma_1}_{T'}(p_5,\theta_2)$ to $T'$. Then, since $(p_5,\theta_2)\xrightarrow{\epsilon}_{T'}(p_5,\theta_2)$ and $r_5\in \theta_2$, Rule $(\alpha_1)$ adds the transition $(p_0,\theta_2)\xrightarrow{\gamma_0}(p_4,\theta_2)$ to $T'$.
		\item Now we have $(p_2,\theta_2)\xrightarrow{\gamma_2}_{T'}(p_5,\theta_2)$ and  $(p_2,\theta_2)\xrightarrow{\gamma_0}_{T'}(p_2,\theta_3)$, since $r_c^4,r_7\in \theta_2$, Rule $\alpha_2$ adds the transitions $(p_4,\theta_0)\xrightarrow{\gamma_2}(p_5,\theta_2)$ and $(p_4,\theta_0)\xrightarrow{\gamma_0}(p_2,\theta_3)$ to $T'$ where $(\theta_2 \setminus \{ r_7\} \cup )\{ r_1\} =\theta_0.$
		\item Since $(p_4,\theta_2)\xrightarrow{\gamma_1}_{T'}(p_5,\theta_2)$ and $r_2,r_c^1\in \theta_2$, Rule $(\alpha_2)$ adds the transition  $(p_5,\theta_3)\xrightarrow{\gamma_1}(p_5,\theta_2)$ to $T'$ where $(\theta_2\setminus \{ r_2\}) \cup \{r_6\}=\theta_3.$
		\item Since  $(p_5,\theta_3)\xrightarrow{\gamma_1}_{T'}(p_5,\theta_2)$ and $r_6\in \theta_3$,  Rule $(\alpha_1)$  adds the transition $(p_2,\theta_3)\xrightarrow{\gamma_2}(p_5,\theta_2)$ to $T'$.
		\item No further additions are possible, so the procedure terminates.
	\end{enumerate}}
\end{example}

\subsection{Proof of Theorem \ref{pre}}

Let us now prove Theorem \ref{pre}. To prove this theorem, we first introduce the following lemma.

\begin{lemma}
\label{lemmapre}
	For every configuration $(\langle p, w \rangle,\theta_0) \in L(A)$, if $(\langle p',w'\rangle,\theta) \Rightarrow_{\mathcal{P}}^* (\langle p, w \rangle,\theta_0)$,  then $(p',\theta)\xrightarrow{w'}_{T'}q$ for some final state $q$ of $\mathcal{A}_{pre^*}$. 
\end{lemma}

\begin{Proof}
\noindent
	Assume  $(\langle p',w'\rangle,\theta) \stackrel{i}{\Rightarrow}_{\mathcal{P}} (\langle p,w\rangle,\theta_0)$. We proceed by induction on $i$.

\medskip
\noindent
 \textbf{Basis.} $i=0$. Then $\theta =\theta_0,p'=p$ and $w=w'$. Since $(\langle p,w \rangle,\theta_0) \in L(A)$, we have $(p,\theta_0)\xrightarrow{w}_{T'} q$ always holds for some final state $q$ i.e.  $(p',\theta)\xrightarrow{w'}_{T'} q$ holds.

\medskip
\noindent
\textbf{Step.} $i>0$. Then there exists a configuration $(\langle p'', u \rangle,\theta'')$ such that 
\begin{equation*}
(\langle p', w'\rangle,\theta) \Rightarrow_{\mathcal{P}} (\langle p'',u\rangle, \theta'') \stackrel{i-1}{\Rightarrow}_{\mathcal{P}} (\langle p, w \rangle,\theta_0)
\end{equation*}
We apply the induction hypothesis to $(\langle p'',u\rangle, \theta'') \stackrel{i-1}{\Rightarrow} (\langle p, w \rangle,\theta_0)$, and obtain $(p'',\theta'')\by{u}_{T'}q$ for $q\in F$.

 Let $w_1,u_1\in \Gamma^*,\gamma'\in \Gamma$ be such that $w'=\gamma'  w_1$, $u=u_1  w_1$. Let $q'$  be a state of $\mathcal{A}_{pre^*}$ s.t.
\begin{equation}
	(p'',\theta''){\by{u_1}}_{T'}q' {\by{w_1}}_{T'}q
\end{equation}
There are two cases depending on which rule is applied to get  $(\langle p', w'\rangle,\theta) \Rightarrow (\langle p'',u\rangle, \theta'')$.
\begin{enumerate}
	\item Case $(\langle p', w'\rangle,\theta) \Rightarrow (\langle p'',u\rangle, \theta'') $ is  obtained by a rule of the form: $\langle p',\gamma' \rangle \hookrightarrow \langle p'',u_1 \rangle \in \Delta$.  In this case, $\theta''=\theta.$
	   By the saturation rule $\alpha_1$, we have 
	  
	  \begin{equation}
	  	(p',\theta'') {\by{\gamma'}}_{T'}q'
	  \end{equation}
	  
	 Putting (1) and (2) together, we can obtain that 
	 \begin{equation}
	 	\pi=(p',\theta'') {\by{\gamma'}}_{T'}q'  {\by{w_1}}_{T'}q
	 \end{equation}
	 
	 Thus, $(p',\theta'')\by{\gamma'w_1}_{T'}q$ i.e. $(p',\theta)\by{w'}q$ for some final state $q\in F$.	  
	  
	 	\item Case $(\langle p', w'\rangle,\theta) \Rightarrow (\langle p'',u\rangle, \theta'') $ is obtained by a rule of the form $p' \smrule{(r_1,r_2)} p'' \in \Delta_c$. I.e $\theta'' \neq \theta.$  In this case, $u_1=\gamma'$. By the saturation rule $\alpha_2$, we obtain that
	 	\begin{equation}
	 		(p',\theta){\by{\gamma'}}_{T'}q' \text{ where }\theta''=\theta \backslash \{ r_1\} \cup \{ r_2\}. 
	 	\end{equation}
	 	
	 	Putting (1) and (4) together, we have the following path
	 	\begin{equation}
	 		(p',\theta){\by{\gamma'}}_{T'} q'{\by{w_1}}_{T'}q. \text{ I.e. } (p',\theta){\by{w'}}_{T'}q \text{ for }q \in F
	 	\end{equation}
	 	\end{enumerate}
\end{Proof}
\begin{lemma}
\label{postlemma}
	 If a path $ \pi=(p,\theta)\xrightarrow{w}_{T'}q$ for $\theta \subseteq \Delta \cup \Delta_c$ is in $\mathcal{A}_{pre^*}$, then
	 \begin{enumerate}
	 	\item[(I)]   $(\langle p,w \rangle, \theta) \Rightarrow^* (\langle p',w' \rangle,\theta_0)$ holds for a configuration $(\langle p',w' \rangle,\theta_0)$ s.t. $(p',\theta_0){\by{w'}}_{T} q$ in the initial $\mathcal{P}$-automaton $\mathcal{A}$;
	 	\item[(II)] Moreover, if $q$ is an initial state i.e. in the form $(p,\theta)$, then $w'=\epsilon$.
 	 \end{enumerate}
\end{lemma}

\begin{Proof}
 Let $\mathcal{A}_{pre^*}=(Q,\Gamma,T,P,F)$ be the  $\mathcal{P}$-automaton computed by the saturation procedure. In this proof, we use $\xlongrightarrow[i]{}_{T'}$ to denote the transition relation of $\mathcal{A}_{pre^*}$ obtained after adding $i$-transitions using the saturation procedure. In particular, since initially  $\mathcal{A}_{pre^*}=\mathcal{A}$,  $\mathcal{A}_{pre^*}$ contains the path $(p',\theta_0)\xlongrightarrow{w'}_{T}q$ where $(\langle p',w' \rangle,\theta_0)\in L(A)$, then we write $(p',\theta_0)\xlongrightarrow[0]{w'}_{T}q$.

\medskip
\noindent
Let $i$ be an index such that  $ \pi=(p,\theta)\xlongrightarrow[i]{w}_{T'}q$  holds. We shall prove (I) by induction on $i$. Statement (II) then follows immediately from the fact that initial states have no incoming transitions in $\mathcal{A}$.  

\medskip
\noindent
\textbf{Basis.} $i=0$. Since $(\langle p,w\rangle,\theta)\Rightarrow^*(\langle p,w\rangle,\theta)$ always holds, take then $ p=p',w=w'$ and $\theta_0=\theta$.

\medskip
\noindent
 \textbf{Step.} $i>0$. Let $t=((p_1,\theta_1), \gamma, q')$ be the $i$-th transition added to $\mathcal{A}_{pre^*}$ and $j$ be the number of times that $t$ is used in the path $ (p,\theta){\xlongrightarrow[i]{w}}_{T'}q$. The proof is by induction on $j$. If $j=0$, then we have $(p,\theta){\xlongrightarrow[i-1]{w}}_{T'}q$ in the automaton, and we apply the induction hypothesis (induction on $i$) then we obtain  $(\langle p,w\rangle, \theta) \Rightarrow^* (\langle p',w' \rangle,\theta_0)$ for a configuration $(\langle p',w' \rangle,\theta_0)$ s.t. $(p',\theta_0){\by{w'}}_{T} q$ in the initial $\mathcal{P}$-automaton $\mathcal{A}$.   So assume that $j>0$. Then, there exist $u$ and $v$ such that $w=u\gamma v$ and 
\begin{equation*}
\tag{1}
(p,\theta) {\xlongrightarrow[i-1]{u}}_{T'} (p_1,\theta_1) {\xlongrightarrow[i]{\gamma}}_{T'}q'{\xlongrightarrow[i]{v}}_{T'}q
\end{equation*}

The application of the induction hypothesis (induction on $i$) to $(p,\theta) \xlongrightarrow[i-1]{u}_{T'} (p_1,\theta_1)$  (notice that $(p_1,\theta_1)$ is an initial state) gives that 
\begin{equation*}
\tag{2}
(\langle p,u\rangle,\theta) \Rightarrow^* (\langle p_1, \epsilon\rangle,\theta_1)
\end{equation*}

There are 2 cases depending on whether transition $t$ was added by saturation rule $\alpha_1$ or $\alpha_2$.
\medskip

\begin{enumerate}
	\item Case $t$ was added  by rule $\alpha_1$: There exist $p_2\in P$ and $w_2\in \Gamma^*$ such that
	\begin{equation*}
	\tag{3}
	r=\langle p_1,\gamma \rangle \hookrightarrow \langle p_2, w_2 \rangle \in \Delta \cap \theta_1
	\end{equation*} 
	and $\mathcal{A}_{pre^*}$ contains the following path:
	\begin{equation*}
	\tag{4}
	\pi'=(p_2,\theta_1)\xlongrightarrow[i-1]{w_2}_{T'}q'\xlongrightarrow[i]{v}_{T'}q
	\end{equation*}

	Applying the transition rule $r$ gets that
	\begin{equation*}
		\tag{5}
		(\langle p_1, \gamma v\rangle,\theta_1) \Rightarrow (\langle p_2, w_2 v\rangle,\theta_1)
	\end{equation*}
	
	By induction on $j$ (since transition $t$ is used $j-1$ times in $\pi'$), we get from (4) that 
	\begin{equation*}
		\tag{6}
		\begin{aligned}
			(\langle p_2,w_2 v\rangle,\theta_1) \Rightarrow^* 	(\langle p',w'\rangle,\theta_0) \text{ s.t. }(p',\theta_0){\by{w'}}_{T'} q \\
			\text{ in the initial }\mathcal{P} \text{-automaton }\mathcal{A}
		\end{aligned}
	\end{equation*}

	Putting (2) ,(5) and (6) together, we can obtain that 
	
	\begin{equation*}
	\begin{aligned}
		(\langle p,w\rangle,\theta)=(\langle p, u \gamma v\rangle,\theta) \Rightarrow^* (\langle p_1, \gamma v\rangle,\theta_1) \Rightarrow \\
		(\langle p_2, w_2 v\rangle,\theta_1)\Rightarrow^* 	(\langle p',w'\rangle,\theta_0) 
	\end{aligned}
	\end{equation*}
 	such that $(p',\theta_0){\by{w'}}_{T} q$ in the initial $\mathcal{P}$-automaton $\mathcal{A}$
 	
 	\medskip
	\item Case $t$ was added by rule $\alpha_2:$ there exist $p_2 \in P$ and $\theta''\subseteq \Delta \cup \Delta_c$ such that
	\begin{equation*}
	\tag{7}
		p_1 \smrule{(r_1,r_2)} p_2 \in \Delta_c \cap \theta'',\theta''=(\theta_1 \backslash \{r_1\}) \cup \{r_2\}
	\end{equation*}
	and the following path in the current automaton ( self-modifying rule won't change the stack) with $r\in\theta'':$	
	\begin{equation*}
	\tag{8}
	(p_2,\theta'') {\xrightarrow[i-1]{\gamma}_{T'}} q' \xlongrightarrow[i]{v}_{T'}q
	\end{equation*}
	Applying the transition rule, we can get from (7) that
	\begin{equation*}
	\tag{9}
	(\langle p_1,\gamma v\rangle, \theta_1) \Rightarrow (\langle p_2,\gamma v \rangle, \theta'')
	\end{equation*}
	We can apply the induction hypothesis (on $j$) to (8), and obtain
	\begin{equation*}
	\tag{10}
	\begin{aligned}
		(\langle p_2,\gamma v\rangle, \theta'') \Rightarrow^* (\langle p',w' \rangle,\theta_0) \text{ s.t. }(p',\theta_0){\by{w'}}_{T} q \\
		\text{ in the initial }\mathcal{P} \text{-automaton }\mathcal{A}
	\end{aligned}
	\end{equation*}
	
	From (2),(9) and (10), we get
	
	\begin{equation*}
	\begin{aligned}
		(\langle p,w\rangle, \theta) =(\langle p,u \gamma v\rangle, \theta)\Rightarrow^* (\langle p_1,\gamma v\rangle,\theta_1)\Rightarrow \\
		(\langle p_2,\gamma v\rangle,\theta'')\Rightarrow^* (\langle p',w' \rangle,\theta_0)
	\end{aligned}
			\end{equation*}
such that $(p',\theta_0){\by{w'}}_{T} q$ in the initial $\mathcal{P}$-automaton $\mathcal{A}$.

\end{enumerate}
\end{Proof}

\noindent
Then, we can prove  Theorem \ref{pre}: 
\medskip

\begin{Proof}
	Let $(\langle p, w\rangle, \theta)$ be a configuration of $pre^*(L(A))$. Then $(\langle p,w\rangle,\theta) \Rightarrow^*(\langle p',w'\rangle, \theta_0)$ for a configuration $(\langle p',w' \rangle, \theta_0)$ s.t. $(p',\theta_0){\by{w'}}_{T'} q$ is a path in $\mathcal{A}$ for $q\in F$. By lemma 1, we can obtain that there exists a path $(p,\theta) \xrightarrow{w}_{T'}q$ for some final state $q$ of $\mathcal{A}_{pre^*}$. So $(\langle p, w \rangle,\theta)$ is recognized by $\mathcal{A}_{pre^*}$.

Conversely, let $(\langle p, w \rangle,\theta)$ be a configuration accepted by $\mathcal{A}_{pre^*}$ i.e. there exists a path $(p,\theta) \xrightarrow{w}_{T'}q$  in $\mathcal{A}_{pre^*}$ for some final state $q\in F$. By lemma 2, there exists a configuration $(\langle p',w' \rangle, \theta_0)$ s.t. there exist a path $(p',\theta_0) \xrightarrow{w'}_{T}q$ in the initial automaton $\mathcal{A}$ and $(\langle p,w \rangle, \theta)\Rightarrow^* (\langle p',w' \rangle, \theta_0) $. Because $q$ is a final state, we have $(\langle p',w' \rangle,\theta_0) \in L(A)$ i.e. $(\langle p,w \rangle, \theta) \in pre^*(L(A))$.

\end{Proof}

\section{Efficient computation of $post^*$ images}
Let $\mathcal{P}=(P,\Gamma,\Delta,\Delta_c)$ be a SM-PDS, and let $\mathcal{A}=(Q,\Gamma,T,P,F)$ be a  $\mathcal{P}$-automaton that represents a regular
set of configurations $\mathcal{C}$ ( $\mathcal{C}=L(\mathcal{A})$). Similarly, it is not optimal  to compute $post^*( \mathcal{C})$ using the translations of 
Sections \ref{section-PDS} and \ref{section-symPDS} to compute  equivalent PDSs or symbolic PDSs, and then apply the algorithms of  \cite{esparza,Schwoon:2007vs}. 
We present in this section a \textbf{ direct} and   \textbf{ efficient}  algorithm that computes $post^*( \mathcal{C})$.
We assume w.l.o.g. that $\mathcal{A}$ has no transitions leading to an initial state. Moreover, we assume that the rules of $\Delta$ are of the form
$\langle p,\gamma \rangle \hookrightarrow \langle p',w \rangle$, where $|w|\le 2$. This is not a  restriction, indeed, a rule of the form 
$\langle p,\gamma \rangle \hookrightarrow \langle p',\gamma_1\cdots\gamma_n \rangle$, $n> 2$ can be replaced by the following rules:
\begin{itemize}
	\item $\langle p,\gamma \rangle \hookrightarrow \langle p_1, a_1\gamma_n \rangle$
	\item $\langle p_1,a_1 \rangle \hookrightarrow \langle p_2, a_2\gamma_{n-1} \rangle$
	\item $\langle p_2,a_2 \rangle \hookrightarrow \langle p_3, a_3\gamma_{n-2} \rangle$
	\item $\cdots$,
	\item $\langle p_{n-2},a_{n-2} \rangle \hookrightarrow \langle  {p'}, \gamma_1\gamma_{2} \rangle$
\end{itemize}

As previously, the construction of $\mathcal{A}_{post^*}$ consists in adding iteratively new transitions to the automaton $\mathcal{A}$ according
to  saturation rules (reflecting the forward application of the transition
rules in the system).
We define $\mathcal{A}_{post^*}$ to be the $\mathcal{P}$-automaton 
$( {Q'},\Gamma,T',P,F)$, where $T'$ is computed using the following saturation rules  {and $Q'$ is the smallest set s.t. $Q\subseteq Q'$  and for every $r=\langle p,\gamma \rangle \hookrightarrow \langle p',\gamma_1 \gamma_2\rangle \in \Delta,  q_{p'\gamma_1}^{\theta} \in Q'$ where $q_{p'\gamma_1}^{\theta}$ is the new state labelled with $p',\gamma_1$ and $\theta$}: 
initially $T'=T$;

\begin{enumerate}
	\item[$\beta_1$:] If $r=\langle p,\gamma \rangle \hookrightarrow \langle p',\epsilon \rangle \in \Delta$ and there exists in $T'$ a path  $\pi=(p,\theta) \by{\gamma}_{T'}q$ with $r\in \theta$, then add $((p',\theta),\epsilon,q)$ to $T'$.
	
	\item[$\beta_2$:] If $r=\langle p,\gamma \rangle \hookrightarrow \langle p',\gamma' \rangle \in \Delta$ and there exists in $T'$ a path  $\pi=(p,\theta) \by{\gamma}_{T'}q$ with $r\in \theta$, then add $((p',\theta),\gamma',q)$ to $T'$.
	
	\item[$\beta_3$:] If $r=\langle p,\gamma \rangle \hookrightarrow \langle p',\gamma_1\gamma_2 \rangle \in \Delta$ and there exists in $T'$ a path  $\pi=(p,\theta) \by{\gamma}_{T'}q$ with $r\in \theta$. 
		Add $t'=((p',\theta),\gamma_1,q_{p'\gamma_1}^{\theta})$ and $t''=(q_{p'\gamma_1}^{\theta},\gamma_2,q)$ to $T'$.
	
	\item[$\beta_4$:]  if $r=p\smrule{(r_1,r_2)} p' \in \Delta_c$ and there exists in $T'$ a path  $\pi=(p,\theta)\by{\gamma}_{T'}q$, where 
	$\gamma\in\Gamma$ with $r\in\theta$,  and $r_1\in\theta$, then add  $t'=((p',\theta'),\gamma,q)$ where $\theta'=(\theta\setminus \{r_1\}) \cup \{r_2\}\}$.
	
\end{enumerate}

The  procedure above  terminates since there is a finite number of states and phases.

Let us explain intuitively the role of the saturation rules above.
Consider a path in the automaton of the form $(p,\theta)\by{\gamma}_{T'}q\by{w'}_{T'}q_F$, where $q_F\in F$. 
This means, by definition of $\mathcal{P}$-automata,  that the configuration $c=(\langle p,\gamma w'\rangle,\theta)$ is accepted by $\mathcal{A}_{post^*}$.

Let $r=\langle p,\gamma \rangle \hookrightarrow \langle p',\epsilon \rangle \in \Delta$.
If $r$ is in $\theta$, then the configuration $c'=(\langle p', w'\rangle,\theta)$ is a successor of $c$. Therefore, it should be added to 
$\mathcal{A}_{post^*}$. This configuration is accepted by the run $(p',\theta)\by{\epsilon}_{T'}q\by{w'}_{T'}q_F$ added by rules  ($\beta_1$).

If $\theta$ contains the rule  $r=\langle p,\gamma \rangle \hookrightarrow \langle p',\gamma' \rangle \in \Delta$,
then the configuration $c'=(\langle p', \gamma' w'\rangle,\theta)$ is a successor of $c$. Therefore, it should be added to 
$\mathcal{A}_{post^*}$. This configuration is accepted by the run $(p',\theta)\by{\gamma'}_{T'}q\by{w'}_{T'}q_F$ added by rules  ($\beta_2$).

If  $r=\langle p,\gamma \rangle \hookrightarrow \langle p',\gamma_1\gamma_2 \rangle \in \Delta$ is in $\theta$,
then the configuration $c'=(\langle p', \gamma_1 \gamma_2 w'\rangle,\theta)$ is a successor of $c$. Therefore, it should be added to 
$\mathcal{A}_{post^*}$. 
This configuration is accepted by the run $(p',\theta)\by{\gamma_1}_{T'}q_{p'\gamma_1}^{\theta}\by{\gamma_2}_{T'}q\by{w'}_{T'}q_F$ added by rules  ($\beta_3$).

Rule ($\beta_4$) deals with modifying rules: Let  $r=p\smrule{(r_1,r_2)} p' \in \Delta_c$. 
If $r$ and $r_1$ are  in $\theta$, then the configuration $c'=(\langle p',\gamma w'\rangle,\theta')$ is a successor of $c$, where  
$\theta'=(\theta\setminus \{r_1 \}) \cup \{r_2\}$. Therefore, it should be added to 
$\mathcal{A}_{post^*}$. This configuration is accepted by the run $(p',\theta')\by{\gamma}_{T'}q\by{w'}_{T'}q_F$ added by rules  ($\beta_4$).

\medskip

Thus, we can show that:
\begin{theorem}
\label{post}
	$\mathcal{A}_{post^*}$  recognizes the set
	$post^*({L(A)})$.
\end{theorem}

Before proving this theorem, let us illustrate the construction on 2 examples.
\begin{figure}[htbp]
\centering
\includegraphics[width=8cm,height=5cm]{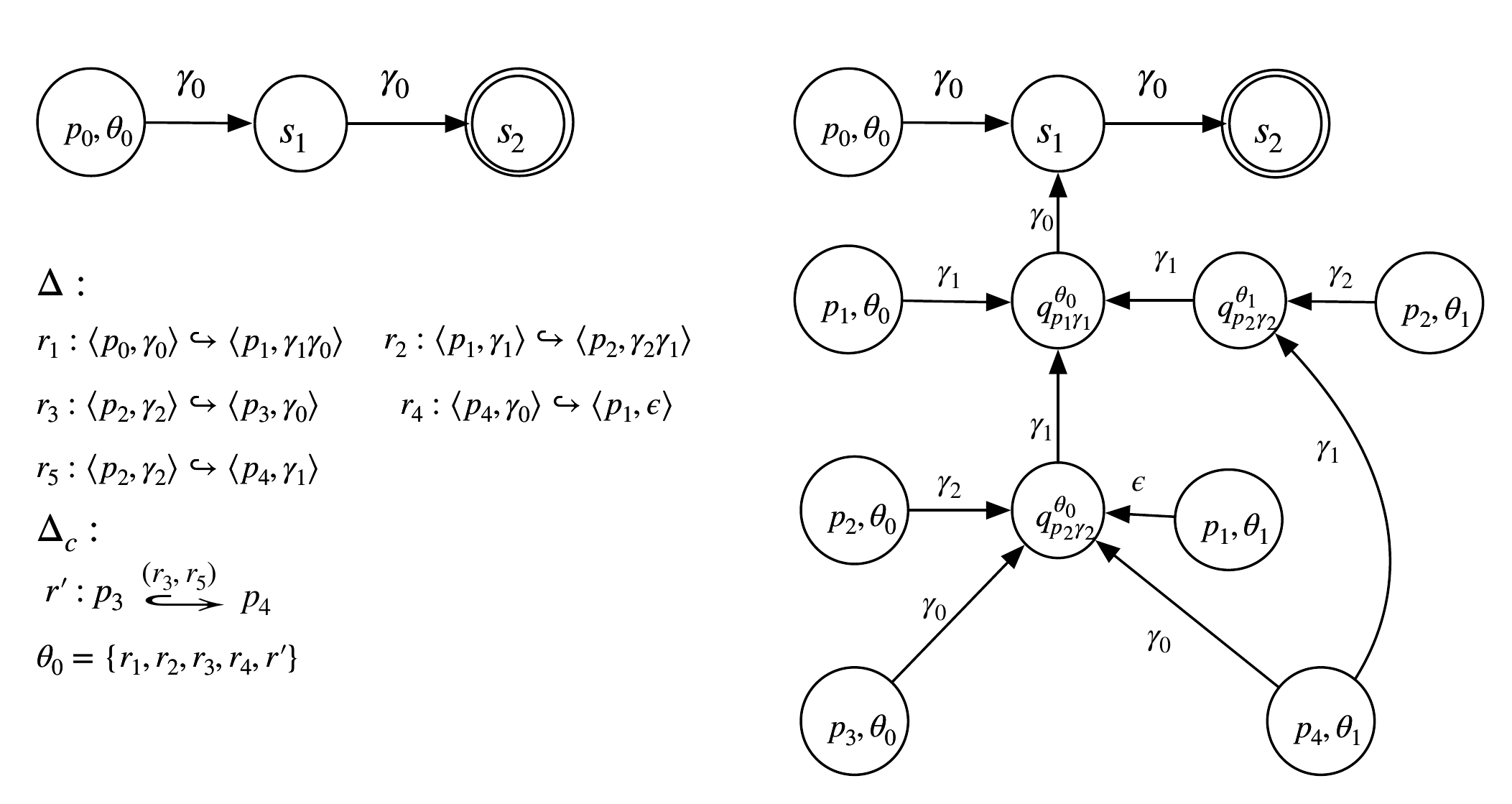}
\caption{The automata $\mathcal{A}$ (left) and $\mathcal{A}_{post^*}$ (right)}\label{fig:postex}
\end{figure}

\begin{example}
 {	Let us illustrate this procedure by an example. Consider  the SM-PDS shown in the left half of Fig. \ref{fig:postex} and the automaton $\mathcal{A}$ from Fig. \ref{fig:postex} that accepts the set $C=\{(\langle p_0,\gamma_0 \gamma_0 \rangle, \theta_0)\}$ where $(p_0,\theta_0)$ is the initial state and $s_2$ is the  {final} state. Then the result $\mathcal{A}_{post^*}$ of the algorithm is shown in the right half of  Fig. \ref{fig:postex}. The result is derived through the following steps:
	\begin{enumerate}
		\item First, since $(p_0,\theta_0)\xrightarrow{\gamma_0}_{T'} s_1$ and  $r_1\in \theta_0$, Rule $(\beta_3)$ generates a new state $q_{p_1\gamma_1}^{\theta_0}$ and adds the two transitions: $(p_1,\theta_0) \xrightarrow{\gamma_1} q_{p_1\gamma_1}^{\theta_0}$ and $q_{p_1\gamma_1}^{\theta_0} \xrightarrow{\gamma_0} s_1$ to $T'$.
		\item Since $(p_1,\theta_0) \xrightarrow{\gamma_1}_{T'} q_{p_1\gamma_1}^{\theta_0}$ and $r_2 \in \theta_0$,   Rule $(\beta_3)$ generates a new state   $q_{p_2\gamma_2}^{\theta_0}$ and adds two transitions : $(p_2,\theta_0) \xrightarrow{\gamma_2} q_{p_2\gamma_2}^{\theta_0}$ and $q_{p_2\gamma_2}^{\theta_0} \xrightarrow{\gamma_1}  q_{p_1\gamma_1}^{\theta_0}$ to $T'$.
		\item Because  $(p_2,\theta_0) \xrightarrow{\gamma_2}_{T'} q_{p_2\gamma_2}^{\theta_0}$ and $r_3 \in \theta_0$, Rule $(\beta_1)$ adds the transition $(p_3,\theta_0) \xrightarrow{\gamma_0}q_{p_2\gamma_2}^{\theta_0}$ to $T'$.
		\item Since $(p_3,\theta_0) \xrightarrow{\gamma_0}_{T'}q_{p_2\gamma_2}^{\theta_0}$ and  $r' \in \theta_0$, Rule $(\beta_4)$ adds the transition $(p_4,\theta_1) \xrightarrow{\gamma_0} q_{p_2\gamma_2}^{\theta_0}$ to $T'$ where $\theta_1 =(\theta_0 \setminus \{r_3\} ) \cup \{r_5\} =\{ r_1,r_2,r_4,r_5,r'\}$.
		\item Since $(p_4,\theta_1) \xrightarrow{\gamma_0}_{T'} q_{p_2\gamma_2}^{\theta_0}$ and $r_4 \in \theta_1$, Rule $(\beta_1)$ adds the transition $(p_1,\theta_1)\xrightarrow{\epsilon}q_{p_2\gamma_2}^{\theta_0}$ to $T'$.
		\item Then, since there is a path  $(p_1,\theta_1)\xrightarrow{\gamma_1}_{T'}^*q_{p_1\gamma_1}^{\theta_0}$ and  $r_2\in \theta_1$, Rule $(\beta_3)$ generates         new state $q_{p_2,\gamma_2}^{\theta_1}$ and adds two transitions $(p_2,\theta_1) \xrightarrow{\gamma_2}q_{p_2,\gamma_2}^{\theta_1} $ and $q_{p_2,\gamma_2}^{\theta_1} \xrightarrow{\gamma_1} q_{p_1\gamma_1}^{\theta_0}$ to $T'$.
		\item Since $(p_2,\theta_1) \xrightarrow{\gamma_2}_{T'}q_{p_2,\gamma_2}^{\theta_1} $  and $r_5 \in \theta_1$, Rule $(\beta_2)$ adds the transition $(p_4,\theta_1)\xrightarrow{\gamma_1}q_{p_2,\gamma_2}^{\theta_1}$ to $T'$.
		\item No unprocessed matches remain. The procedure terminates.
	\end{enumerate}}
\end{example}
\begin{figure*}[htbp]
\centering
\includegraphics[width=14cm,height=7cm]{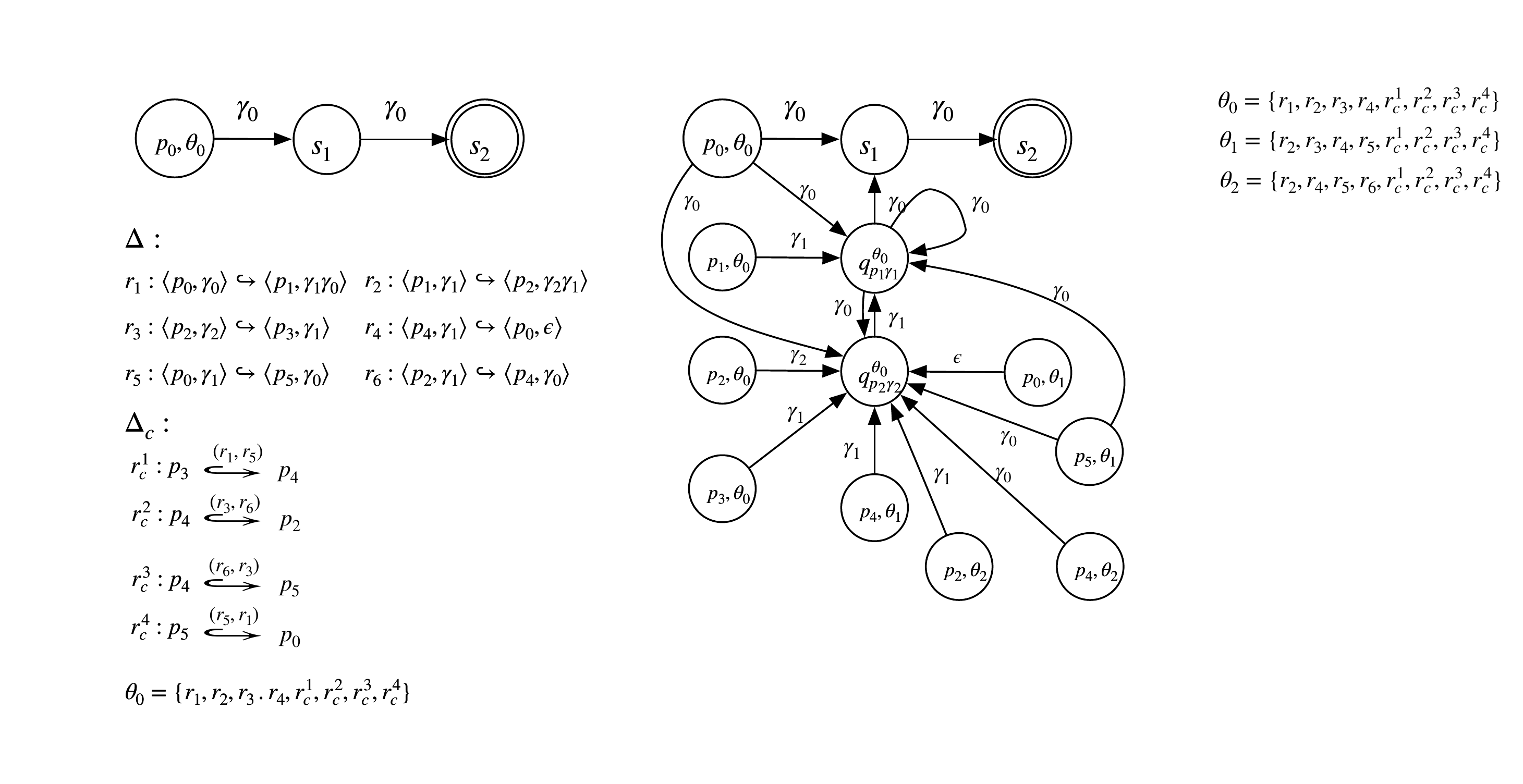}
\caption{The automata $\mathcal{A}$ (left) and $\mathcal{A}_{post^*}$ (right)}\label{fig:compost}
\end{figure*}
\begin{example}
 {
	Let us illustrate this procedure by another example. Consider the SM-PDS shown in the left half of Fig. \ref{fig:compost} where $(p_0,\theta_0)$ is the initial state and $s_2$ is the  {final} state. The result $\mathcal{A}_{post^*}$ of the algorithm is shown in the right half of Fig. \ref{fig:compost} obtained as follows:
	\begin{enumerate}
		\item First, since $(p_0,\theta_0)\xrightarrow{\gamma_0}_{T'}s_1$ and $r_1\in \theta_0$, Rule $(\beta_3)$ generates a new state $q_{p_1\gamma_1}^{\theta_0}$ and adds two transitions: $(p_1,\theta_0) \xrightarrow{\gamma_1} q_{p_1\gamma_1}^{\theta_0}$ and $q_{p_1\gamma_1}^{\theta_0} \xrightarrow{\gamma_0} s_1$ to $T'$.
		\item  Since $(p_1,\theta_0) \xrightarrow{\gamma_1}_{T'} q_{p_1\gamma_1}^{\theta_0}$ and  $r_2 \in \theta_0$,   Rule $(\beta_3)$ generates a new state $q_{p_2\gamma_2}^{\theta_0}$ and adds two transitions: $(p_2,\theta_0) \xrightarrow{\gamma_2} q_{p_2\gamma_2}^{\theta_0}$ and $q_{p_2\gamma_2}^{\theta_0} \xrightarrow{\gamma_1}  q_{p_1\gamma_1}^{\theta_0}$ to $T'$.
		\item Because $(p_2,\theta_0) \xrightarrow{\gamma_2}_{T'} q_{p_2\gamma_2}^{\theta_0}$ and $r_3\in \theta_0$, Rule $(\beta_2)$ adds $(p_3,\theta_0) \xrightarrow{\gamma_1} q_{p_2\gamma_2}^{\theta_0}$ to $T'$. 
		\item  Since $(p_3,\theta_0) \xrightarrow{\gamma_1}_{T'} q_{p_2\gamma_2}^{\theta_0}$  and $r_c^1,r_1\in \theta_0$, Rule $(\beta_4)$ adds the transition $(p_4,\theta_1)\xrightarrow{\gamma_1}q_{p_2\gamma_2}^{\theta_0}$ to $T'$ where $\theta_1=(\theta_0 \setminus \{r_1\}) \cup \{r_5 \}$.
		\item Since $(p_4,\theta_1)\xrightarrow{\gamma_1}_{T'}q_{p_2\gamma_2}^{\theta_0}$ and $r_4 \in \theta_1$, Rule $(\beta_1)$ adds the transition $(p_0,\theta_1)\xrightarrow{\epsilon}q_{p_2\gamma_2}^{\theta_0}$ to $	T'$. Then there is a   path $(p_0,\theta_1)\xrightarrow{\gamma_1}_{T'}^* q_{p_1\gamma_1}^{\theta_0}$, since $r_5\in \theta_1$, Rule $(\beta_2)$ adds the transition  $(p_5,\theta_1)\xrightarrow{\gamma_0}q_{p_1\gamma_1}^{\theta_0}$ to $T'$.
		\item  Since  $(p_5,\theta_1)\xrightarrow{\gamma_0}_{T'}q_{p_1\gamma_1}^{\theta_0}$ and $r_c^4,r_5\in \theta_1$, Rule $(\beta_4)$ adds the transition $(p_0,\theta_0)\xrightarrow{\gamma_0}q_{p_1\gamma_1}^{\theta_0}$ to $T'$ where $(\theta_1 \setminus\{r_5\} )\cup \{ r_1\} =\theta_0$.
		\item Since $(p_0,\theta_0)\xrightarrow{\gamma_0}_{T'}q_{p_1\gamma_1}^{\theta_0}$ and  $r_1\in \theta_0$, Rule $(\beta_3)$ adds the transitions  $(p_1,\theta_0) \xrightarrow{\gamma_1} q_{p_1\gamma_1}^{\theta_0}$ and $q_{p_1\gamma_1}^{\theta_0} \xrightarrow{\gamma_0} q_{p_1\gamma_1}^{\theta_0}$ to $T'$.
		\item Because $(p_4,\theta_1)\xrightarrow{\gamma_1}_{T'}q_{p_2\gamma_2}^{\theta_0}$ and $r_c^2\in \theta_1$, Rule $(\beta_4)$ adds the transition $(p_2,\theta_2)\xrightarrow{\gamma_1}q_{p_2\gamma_2}^{\theta_0}$ to $T'$.
		\item Since $(p_2,\theta_2)\xrightarrow{\gamma_1}_{T'}q_{p_2\gamma_2}^{\theta_0}$ and $r_6\in \theta_2$, Rule $(\beta_{2})$ adds the transition $(p_4,\theta_2)\xrightarrow{\gamma_0}q_{p_2\gamma_2}^{\theta_0}$ to $T'$.
		\item Since $p_4,\theta_2\xrightarrow{\gamma_0}_{T'}q_{p_2\gamma_2}^{\theta_0}$ holds and $r_6,r_c^3\in \theta_2$,  Rule $(\beta_4)$  adds the transition $(p_5,\theta_1) \xrightarrow{\gamma_0}q_{p_2\gamma_2}^{\theta_0}$ to $T'$.
		\item  Then, since $(p_5,\theta_1) \xrightarrow{\gamma_0}_{T'}q_{p_2\gamma_2}^{\theta_0}$ and  $r_c^4\in \theta_1$, Rule $(\beta_4)$  adds the transition  $(p_0,\theta_0)\xrightarrow{\gamma_0}q_{p_2\gamma_2}^{\theta_0}$ to $T'$.
		\item Since $r_1\in \theta_0$ and $(p_0,\theta_0)\xrightarrow{\gamma_0}_{T'}q_{p_2\gamma_2}^{\theta_0}$, Rule ($\beta_3$) adds two transitions: $(p_1,\theta_0) \xrightarrow{\gamma_1} q_{p_1\gamma_1}^{\theta_0}$ and $q_{p_1\gamma_1}^{\theta_0} \xrightarrow{\gamma_0}q_{p_2\gamma_2}^{\theta_0}$ to $T'$.
		\item No more rules can be applied. Thus, the procedure terminates.
	\end{enumerate}}
\end{example}

\subsection{Proof of Theorem \ref{post}}
Let us now prove Theorem \ref{post}. To prove this theorem, we first show the following lemma:

\begin{lemma}
\label{lemmaPost}
	For every configuration $(\langle p,w \rangle,\theta_0) \in L(A)$, if $(\langle p,w \rangle,\theta_0) \Rightarrow^* (\langle p',w'\rangle,\theta) $ then we have a path $\pi=(p',\theta)\xlongrightarrow{w'}_{T'}q$ for some final state $q$ of $\mathcal{A}_{post^*}$. 
\end{lemma}
 \begin{Proof}
 
 Let $i$ be the index s.t. $(\langle p,w\rangle,\theta_0) \stackrel{i}{\Rightarrow}  (\langle p' ,w'\rangle,\theta) $ holds. We  proceed  by induction on $i$.
 
 \medskip
 \noindent
 \textbf{Basis.} $i=0$. Then $p'=p$, $w=w'$ and $\theta_0=\theta$. Since $(\langle p,w\rangle,\theta_0) \in L(A)$, we have $(p,\theta_0)\xlongrightarrow{w}_{T'}q$ for some final state $q$ that implies  $\pi=(p',\theta)\xlongrightarrow{w'}_{T'}q$ is a path of $\mathcal{A}_{post^*}$.
 
 \medskip
 \noindent
\textbf{Step.} $i>0.$ Then there exists a configuration $(\langle p'',u  \rangle,\theta'')$ with
\begin{equation*}
(\langle p,w \rangle,\theta_0)\stackrel{i-1}{\Rightarrow}(\langle p'',u  \rangle,\theta'')\Rightarrow(\langle p',w' \rangle,\theta) 
\end{equation*}
By applying the induction hypothesis (induction on $i$), we can get  that

\begin{equation*}
\tag{1}
(p'',\theta'')\xlongrightarrow{u}_{T'} q \text{ for some } q\in F
\end{equation*} 

Then, let $\gamma\in \Gamma$, $u_1,w_1\in \Gamma^*$ be such that  $u=\gamma u_1$, $w'=w_1 u_1$. Let  $q_1$ be a state of $\mathcal{A}_{post^*}$ s.t. we have the following path in $\mathcal{A}_{post^*}$:

\begin{equation*}
\tag{2}
(p'',\theta'')\xrightarrow{\gamma}_{T'} q_1 {\by{u_1}}_{T'} q 
\end{equation*} 
\medskip
\noindent
There are two cases depending on whether $(\langle p'',u \rangle,\theta'')\Rightarrow(\langle p',w'\rangle,\theta)  $ is corresponding to a self-modifying transition (i.e. ($\theta''=\theta$)) or not.

\begin{enumerate}
	\item Case: $\theta''=\theta$. Then there exists  
	a transition rule   $r:\langle p'',\gamma\rangle \hookrightarrow\langle p',w_1\rangle  \in \Delta$ s.t. $r\in \theta$. There are three possible cases depending on  the length of $w_1:$ 
	\begin{enumerate}
		\item[-] Case $|w_1|=0$ i.e. $w_1=\epsilon$, by applying the saturation rule $\beta_1$,    we can get  
\begin{equation*}
	\tag{3}
	(p',\theta)\xrightarrow{\epsilon}_{T'}q_1
\end{equation*}
	  
Putting (2) and (3) together, we can have $(p',\theta)\xrightarrow{\epsilon}_{T'}q_1 \by{u_1}_{T'} q$ i.e. $(p',\theta)\xrightarrow{w'}_{T'} q$ for some final state $q$ of $\mathcal{A}_{post^*}$.
 
		\item[-] Case $|w_1|=1,$ then let $\gamma'\in \Gamma $ s.t. $w_1=\gamma'$.  {By} applying the saturation rule  $\alpha_2$,    we can get  
\begin{equation*}
	\tag{4}
	(p',\theta)\xrightarrow{\gamma'}_{T'}q_1 
\end{equation*}
	  
Putting (2) and (4) together, we can have $(p',\theta)\xrightarrow{\gamma'}_{T'}q_1 \by{u_1}_{T'} q$ i.e. $(p',\theta)\xrightarrow{w'}_{T'} q$  for some final state $q$ of $\mathcal{A}_{post^*}$.

	\item[-] Case $|w_1|=2,$ let $\gamma_0', \gamma_1'\in \Gamma$ be such that $w_1=\gamma_0'\gamma_1'$. By applying the saturation rule $\alpha_3$,    we can get 
	\begin{equation*}
	\tag{5}
	(p',\theta){\by{\gamma_0'}}_{T'}q^{\theta}_{p'\gamma_0'}{\by{\gamma_1'}}_{T'}q_1	\end{equation*} 
	Putting (2) and (5) together, then we have a path $(p',\theta)\xrightarrow{\gamma_0'}_{T'}q^{\theta}_{p'\gamma_0'}\xrightarrow{\gamma_1'}_{T'}q_1\xrightarrow{u_1}_{T'}q$ i.e. $(p',\theta)\xrightarrow{w'}_{T'} q$  for some final state $q$ of $\mathcal{A}_{post^*}$.
		
	\end{enumerate}
		
	\item Case $\theta'' \neq \theta$. Then there exists a self-modifying transition rule  s.t. $r:p''\smrule{(r_1,r_2)} p'\in \Delta_c \cap \theta''$ and $\gamma=w_1$ and $\theta = (\theta'' \backslash \{r_1\}) \cup \{r_2\}$.
	
	 By applying  rule $\beta_4$ to (2), we have the following path in the automaton:
	
	\begin{equation*}
	\tag{5}
	(p',\theta)\xrightarrow{\gamma}_{T'}q_1\xrightarrow{u_1}_{T'}q 
		\end{equation*}
	i.e.  $(p',\theta)\xrightarrow{w'}_{T'}q$ for some final state $q$ of $\mathcal{A}_{post^*}$.
\end{enumerate}

\end{Proof}
\begin{lemma}
\label{lemmapost2}
	If a path  $ \pi=(p,\theta)\xrightarrow{w}_{T'}q$ is in $\mathcal{A}_{post^*}$, then the following holds:
\begin{enumerate}
	
	\item[(I)] if $q$ is a  state of $\mathcal{A}$, then
	$(\langle p',w'\rangle ,\theta_0)\Rightarrow^* (\langle p,w \rangle,\theta)$ for a configuration
	$(\langle p',w'\rangle,\theta_0)$ such that $(p',\theta_0)\xrightarrow{w'}_{T}q$ is a path in the initial $\mathcal{P}$-automaton $\mathcal{A}$ ;
	
	\item[(II)] if $q$ is a new state of the form $q=q_{p_1\gamma_1}^{\theta_1}$, then
	$(\langle p_1,\gamma_1 \rangle, \theta_1) \Rightarrow^* (\langle p,w \rangle,\theta)$.
	
\end{enumerate}
\end{lemma}
 \begin{Proof} Let $\mathcal{A}_{post^*}=( {Q'},\Gamma, {T'},P,F)$ be the $\mathcal{P}$-automaton computed by the saturation procedure. In this proof, we use $\xlongrightarrow[i]{}_{T'}$ to denote the transition relation $\rightarrow_{T'}$ of $\mathcal{A}_{post^*}$ obtained after adding $i$ transitions using the saturation procedure.

\bigskip
\noindent
Let $i$ be an index such that
$(p,\theta)\xlongrightarrow[i]{w}_{T'}q$ holds. We prove both parts of the lemma  by induction on $i$.

\medskip
\noindent
 \textbf{Basis.} $i=0$.  Only (I) applies.
 Thus, $p'=p$, $\theta_0=\theta$ and $w=w'$. $(\langle p',w'\rangle,\theta) \Rightarrow^* (\langle p',w'\rangle,\theta)$ always holds.
 
 \medskip
 \noindent
\textbf{Step.} $i>1$. Let $t$ be the $i$-th transition added to the automaton.  Let $j$ be the number of times that
$t$ is used in $(p,\theta)\xlongrightarrow[i]{w}_{T'}q$.   
$\mathcal{A}$ has no transitions leading to initial states, and the algorithm does not add any such transitions; therefore, if $t$ starts in an initial state, $t$ can only be used at the start of the path.

The proof is by induction on $j$. If $j=0$, then we have $(p,\theta)\xrightarrow[i-1]{w}_{T'}q $. We apply the induction hypothesis (induction on $i$) then we obtain that there exists a configuration $(\langle p',w'\rangle,\theta_0)$  s.t. $(\langle p',w' \rangle,\theta_0) \Rightarrow^* (\langle p,w \rangle,\theta)$ and $(p',\theta_0)\xrightarrow{w'}_{T} q$ is a path of initial $\mathcal{P}$-automaton $\mathcal{A}$. So
assume that $j>0$. We distinguish three possible cases:
\begin{enumerate}
	\item If $t$ was added by the rule $\beta_1$, $\beta_2$ or  $\beta_3$, then $t=((p_1,\theta_1),v,q_1)$, where $v=\epsilon$ or $v=\gamma_1$. Then, necessarily,  $j=1$ and there exists the following path in the current automaton:
	\begin{equation*}
	\tag{1}
	(p,\theta)=(p_1,\theta_1)\xlongrightarrow[i]{v}_{T'} q_1 \xlongrightarrow[i-1]{w_1}_{T'}q
	\end{equation*}
		
	There are 2 cases depending on whether transition $t$ was added by rule $\beta_4$ or not.
	\begin{enumerate}
		\item[-] Case $t$ was added by rule $\beta_4$:  there exists  a self-modifying transition rule such that $r=p_2 \smrule{(r_1,r_2)}p_1\in\Delta_c$, and  there exists the following path in the current automaton:

	
		\begin{equation*}
		\tag{2}
		(p_2,\theta_2)\xlongrightarrow[i-1]{v}_{T'} q_1 \xlongrightarrow[i-1]{w_1}_{T'}q, \theta_1=\theta_2\backslash \{r_1\} \cup \{r_2\}
				\end{equation*}

		By induction on $(i)$, we get from (2)  that there exists a configuration $ (\langle p',w' \rangle,\theta_0)$ s.t. $(p',\theta_0) \xrightarrow{w'}_{T} q$ is a path in the initial $\mathcal{P}$-automaton $\mathcal{A}$:
		\begin{equation*}
		\tag{3}
   (\langle p',w' \rangle,\theta_0) \Rightarrow^* (\langle p_2, vw_1\rangle,\theta_2) 		
		\end{equation*}
		  By applying the rule  $p_2 \smrule{(r_1,r_2)}p_1$, we get that 
		\begin{equation*}
		\tag{4}
		(\langle p_2, vw_1 \rangle,\theta_2)\Rightarrow(\langle p_1, vw_1 \rangle,\theta_1)
		\end{equation*}
		Thus, putting (3) and (4) together, we get that  there exists a configuration $ (\langle p',w' \rangle,\theta_0)$ s.t. $(p',\theta_0) \xrightarrow{w'}_{T} q$ is a path in the initial $\mathcal{P}$-automaton $\mathcal{A}$  {and}:
		\begin{equation*}
		\tag{5}	
		\begin{aligned}
			 (\langle p',w' \rangle,\theta_0)\Rightarrow^* (\langle p_2,vw_1\rangle,\theta_2)\Rightarrow \\
		(\langle p_1, w \rangle ,\theta_1)=(\langle p,w\rangle,\theta)
		\end{aligned}	
		\end{equation*}

		\item[-] Case $t$ is added by $\beta_1$ or $\beta_2$:  then there exists $p_2\in P$, $\gamma_2\in \Gamma$ such that 
		\begin{equation*}
		\tag{6}	
	r=\langle p_2,\gamma_2\rangle \hookrightarrow \langle p_1,v \rangle\in \Delta
		\end{equation*}
		and $\mathcal{A}_{post^*}$ contains the following path:

		\begin{equation*}
		\tag{7}
		(p_2,\theta_1)\xlongrightarrow[i-1]{\gamma_2}_{T'}q_1\xlongrightarrow[i-1]{w_1}_{T'}q
		\end{equation*}
		By induction on $(i)$,  We can get from (7)  that  there exists a configuration $ (\langle p',w' \rangle,\theta_0)$ s.t. $(p',\theta_0) \xrightarrow{w'}_{T} q$ is a path in the initial $\mathcal{P}$-automaton $\mathcal{A}$  {and}:
		\begin{equation*}
		\tag{8} 
			 (\langle p',w' \rangle,\theta_0)\Rightarrow^* (\langle p_2, \gamma_2 w_1 \rangle,\theta_1)   
		\end{equation*}

		Thus, putting (6) and (8) together, we have that  there exists a configuration $ (\langle p',w' \rangle,\theta_0)$ s.t. $(p',\theta_0) \xrightarrow{w'}_{T} q$ is a path in the initial $\mathcal{P}$-automaton $\mathcal{A}$  {and}:
		\begin{equation*}
		\tag{9}
		\begin{aligned}
			(\langle p',w' \rangle,\theta_0)\Rightarrow^* (\langle p_2,\gamma_2 w_1 \rangle,\theta_1)\Rightarrow \\
			(\langle p_1,w \rangle ,\theta_1)=(\langle p,w\rangle,\theta)	
		\end{aligned}		
		\end{equation*}	
	\end{enumerate}
	
		\medskip
	
	\item If $t$ is the first transition added by rule $\beta_3$ i.e. $t$ is in the form of $((p_1,\theta''),\gamma_1,q_{p_1\gamma_1}^{\theta_1})$. If this transition is new, then there are no transitions outgoing from  $ q_{p_1\gamma_1}^{\theta_1}$. So the only path using $t$ is $(p_1,\theta'') \xlongrightarrow[i]{\gamma_1}_{T'}q_{p_1\gamma_1}^{\theta_1}$. For this path, we only need to prove part (II), and $(\langle p_1,\gamma_1 \rangle, \theta_1) \Rightarrow^*(\langle p_1,\gamma_1 \rangle, \theta_1)$ holds trivially.

\item Let $t=(q_{p_1\gamma_1}^{\theta_1},\gamma'',q')$ be the second transition added by saturation rule $\beta_3$. Then there exist $u$, $v\in \Gamma^*$ s.t. $w=u\gamma''v$ and  the current automaton contains the following path: 
\begin{equation*}
	\tag{10}
	(p,\theta) \xlongrightarrow[i-1]{u}_{T'} q_{p_1\gamma_1}^{\theta_1} \xlongrightarrow[i]{\gamma''}_{T'} q' \xlongrightarrow[i]{v}_{T'} q
\end{equation*}

Because $t$ was added via the saturation rule, then there exist $p_2 \in P$, $\gamma_2\in \Gamma$ and a rule of the form
	\begin{equation*}
	\tag{11}
	\langle p_2,\gamma_2 \rangle \hookrightarrow \langle p_1,\gamma_1 \gamma'' \rangle \in \Delta  \cap \theta_1
	\end{equation*}
and  $\mathcal{A}_{post^*}$ contains the following path:

	\begin{equation*}
		\tag{12}
		(p_2,\theta_1)\xlongrightarrow[i-1]{\gamma_2}_{T'}q'\xlongrightarrow[i]{v}_{T'}q
	\end{equation*}

We apply the induction hypothesis on $i$ and obtain that

\begin{equation*}
	\tag{13}
		(\langle p_1,\gamma_1\rangle,\theta_1)\Rightarrow^* (\langle p,u\rangle,\theta) 	\end{equation*}

We apply the induction hypothesis on $i$ to obtain that there exists a configuration $(\langle p',w'\rangle,\theta_0)$ s.t.  $(p',\theta_0) \xrightarrow{w'}_{T} q$ is a path in the initial $\mathcal{P}$-automaton $\mathcal{A}$  {and}:
\begin{equation*}
	\tag{14}
		(\langle p',w' \rangle,\theta_0)\Rightarrow^* (\langle p_2,\gamma_2 v \rangle,\theta_1) 
			\end{equation*}
	
	Thus, putting (11) (13) and (14) together, we have that there exists a configuration $(\langle p',w'\rangle,\theta_0)$ s.t.  $(p',\theta_0) \xrightarrow{w'}_{T} q$ is a path in the initial $\mathcal{P}$-automaton $\mathcal{A}$  {and}:
	
	\begin{equation*}
	\tag{15}
	\begin{aligned}
		(\langle p',w' \rangle,\theta_0) \Rightarrow^* (\langle p_2,\gamma_2v \rangle,\theta_1) \Rightarrow \\
		(\langle p_1,\gamma_1\gamma'' v\rangle,\theta_1)
 \Rightarrow^* (\langle p, u \gamma'' v \rangle,\theta)=(\langle p, w \rangle,\theta)
	\end{aligned}
	\end{equation*}
\end{enumerate}
\end{Proof}

\noindent
Then we continue to prove Theorem \ref{post}:
\medskip

\begin{Proof}
	Let $(\langle p',w' \rangle, \theta)$ be a configuration of $post^*(L(\mathcal{A}))$. Then there exists a configuration $(\langle p,w \rangle,\theta_0)$ such that there exists a path $(p,\theta_0)\xrightarrow{w}_{T}q$ in the initial automaton $\mathcal{A}$ and $(\langle p,w \rangle,\theta_0) \Rightarrow^* (\langle p', w'\rangle, \theta)$. By Lemma 3, we can have $(p',\theta) \xrightarrow{w'}_{T'} q$ for $q$ is a final state of $\mathcal{A}_{post^*}$. So $(\langle p' ,w'\rangle,\theta)$ is recognized by  $\mathcal{A}_{post^*}$.

\medskip
\noindent
Conversely, let $(\langle p',w' \rangle,\theta)$ be a configuration recognized by
$\mathcal{A}_{post^*}$. Then there exists a path $(p',\theta) \xrightarrow{w'}_{T'}q$  in $\mathcal{A}_{post^*}$ for some final
state $q$. By Lemma 4, since $q$ is a final state, we have
$(\langle p,w \rangle,\theta_0) \Rightarrow_{\mathcal{P}}^* (\langle p',w' \rangle,\theta)$ s.t. there exists a configuration $(\langle p,w \rangle,\theta_0)$ s.t. $(p,\theta_0)\xrightarrow{w}_{T'}q$ is a path in the initial automaton $\mathcal{A}$ i.e.  $(\langle p,w \rangle,\theta_0)\in L(A)$. Therefore, $(\langle p',w'  \rangle,\theta)\in
post^*(L(A))$

\end{Proof}

\section{Experiments}

\subsection{Our algorithms vs. standard $pre^*$ and $post^*$ algorithms of PDSs}
\begin{table*}[htbp]
	\centering
	\begin{tabular}{|c|c|c|c|c|c|c|c|}
		\hline
		\hline
		$|\Delta|+|\Delta_c|$& SM-PDS & PDS &Result1 & Total1 & Symbolic PDS &Result2 & Total2 \\
		\hline
		$10+3$&\textbf{0.08s \& 2MB} & 0.15s \& 3MB&0.00s&0.15s&0.10s \& 2MB&0.00s&0.10s\\
		\hline
		$13+3$ &\textbf{0.10s \& 2MB} & 0.15s \& 3MB&0.00s&0.15s&0.10s \& 2MB&0.00s&0.10s\\
		\hline
		$13+3$&\textbf{0.12s \& 2MB}& 0.15s \& 3MB&0.00s&0.15s& 0.10s \& 2MB&0.00s&0.10s\\
		\hline
		$43+7$&\textbf{0.24s \&3MB}& 3.44s \&4MB&0.02s&3.46s&4.80s \&5MB&0.01s&4.81s\\
		\hline
		$ 110+10$&\textbf{0.38s \&7MB}& 5.15s \&6MB&0.01s&5.16s& 2.71s \&8MB&0.00s&2.71s\\
		\hline
		$120+10$& \textbf{0.42s \&11MB}& 5.20s \&15MB&0.01s&5.21s& 2.79s \&10MB&0.01s&2.80s\\
		\hline
		$255+8$&\textbf{0.65s \& 15M}B&295.41s \& 86MB&0.05s&295.46s&21.41s \& 76MB&0.02s&21.43s\\
		\hline
		$1009+10$ &\textbf{1.49s \&97MB} &11504.2s \& 117MB &2.46s&11506.66s& 14.10s \&471MB&1.74s&15.84s\\
		\hline
		$1899+7$ &\textbf{2.98s \& 210MB} &6538s \& 171MB &4.09s&6542.09s& 124.10s \& 558MB&2.71s&173.71s\\
		\hline
		$2059+8$ &\textbf{3.82s \&423MB} &19525.1s \&113MB &4.19s&19529.29s&20.70s \&713MB&error&-\\
		\hline
		$2099+8$ &\textbf{4.05s \& 32MB }&19031s \& 192MB &4.19s&19035.19s& 124.12s \& 757MB&error&-\\
		\hline
		$2099+9$ &\textbf{7.08s \& 252MB}&29742s \& 198MB &4.28s&29746.28s& 128.12s \& 760MB&error&-\\
		\hline
		$3060+9$ &\textbf{11.36s \& 282MB }& 29993.05s \& 241MB &18.72s&30011.77s&261.07s \& 610MB&error&-\\
		\hline
		$3160+9$ &\textbf{11.99s \& 285MB }& 29252.05s \& 257MB &26.15s&29278.2s&162.55s \& 611MB&error&-\\
		\hline
		$4058+7$ &\textbf{18.06s \& 332MB} & 81408.51s \&307MB &92.68s&81501.19s&802.07s \&1013MB&error&-\\
		\hline
		$4058+8$ &\textbf{19.42s \& 397MB} &82812.51s \&399MB &91.91s&82904.42s&899.07s \& 1020MB&error&-\\
		\hline
		$4158+8$ &\textbf{21.68s \&491MB }&83112.51s \&401MB&97.68s&83210.19&899.19s \&1021MB&error&-\\
		\hline
		$5050+8$ &\textbf{23.26s \&499MB} &93912.51s \&298MB &118.12&94030.63s&205.12s \&375MB&error&-\\
		\hline
	\end{tabular}
	\caption{Our direct  $pre^*$  algorithm vs. standard $pre^*$ algorithms of PDSs}
	\label{pretable}
\end{table*}
We implemented our  algorithms in a tool. To compare the performance of our algorithms against  the approach that consists in translating the SM-PDS into an 
equivalent PDS or symbolic PDS and then apply the standard $post^*$ and $pre^*$ algorithms for PDSs and symbolic PDSs \cite{esparza,Schwoon:2007vs}, we first applied our tool on  randomly
generated  SM-PDSs of various sizes. The results of the comparision using the $pre^*$ (resp. $post^*$) algorithms are  reported in Table 1 (resp. Table 2).

\begin{table*}[htbp]
	\centering
	\begin{tabular}{|c|c|c|c|c|c|c|c|}
		\hline
		\hline
		$|\Delta|+|\Delta_c|$& SM-PDS & PDS &Result1 & Total1 & Symbolic PDS &Result2 & Total2 \\
		\hline
		$10+3$&\textbf{0.12s \& 2MB}& 0.15s \& 3MB&0.00s&0.15s&0.10s \&2MB&0.00s&0.10s\\
		\hline
		$13+3$&\textbf{0.12s \& 2MB}& 0.15s \&3MB&0.00s&0.15s& 0.10s \&2MB&0.00s&0.10s\\
		\hline
		$43+7$&\textbf{0.28s \& 2MB}& 3.44s \& 4MB&0.04s&3.48s& 4.80s \& 5MB&0.02s&4.82s\\
		\hline
		$110+10$&\textbf{0.36s \& 8MB}&5.15s \&6MB&0.01s&5.16s& 2.71s \&8MB&0.00s&2.71s\\
		\hline
		$120+10$&\textbf{0.39s \& 13MB}& 5.20s \&15MB&0.01s&5.21s& 2.79s \& 10MB&0.01s&2.80s\\
		\hline
		$255+8$&\textbf{0.44s \& 15MB}&295.41s \& 86MB&0.05s&295.46s& 21.41s \& 76MB&0.02s&21.43s\\
		\hline
		$1009+10$ &\textbf{1.48s \& 97MB}&11504.2s \& 117MB &2.56s&11506.76s& 14.10s \& 471MB&1.75s&15.85s\\
		\hline
		$1899+7$ &\textbf{3.47s \& 212MB}&6538s \& 171MB &3.89s&6541.89s& 124.10s \& 558MB&2.71s&126.81s\\
		\hline
		$2059+8$ &\textbf{4.03s \&323MB} &19525.1s \& 113MB &3.99s&19528.99s&20.70s \&713MB&error&-\\
		\hline
		$2099+8$ &\textbf{4.15s \&332MB} &19031s \&192MB &3.99s&19034.99s& 124.12s \& 757MB&error&-\\
		\hline
		$2099+9$ &\textbf{4.95s \& 352MB} &29742s \& 198MB &4.18s&29746.18s& 128.12s \& 760MB&error&-\\
		\hline
		$3060+9$ &\textbf{5.71s \& 388MB }&29993.05s \& 241MB &18.12s&30011.17s&261.07s \&610MB&error&-\\
		\hline
		$3160+9$ &\textbf{5.79s \& 415MB }&29252.05s \& 257MB &26.10s&29278.15s&162.55s \& 611MB&error&-\\
		\hline
		$4058+7$ &\textbf{7.56s \& 364MB }&81408.51s \& 307MB &91.68s&81500.19s&802.07s \& 1013MB&error&-\\
		\hline
		$4058+8$ &\textbf{9.76s \& 387MB} &82812.51s \& 399MB &91.71s&82904.22s&899.07s \& 1020MB&error&-\\
		\hline
		$4158+8$ &\textbf{11.85s \& 487MB }&83112.51s \& 401MB &97.28s&83209.79s&899.19s \& 1021MB&error&-\\
		\hline
		$5050+8$ &\textbf{13.04s \& 498MB }&93912.51s \&498MB &112.53s&94025.04s&205.12s \& 375MB&error&-\\
		\hline
	\end{tabular}
	\caption{Our direct  $post^*$  algorithm vs. standard $post^*$ algorithms of PDSs}
	\label{posttable}
\end{table*}

In Table \ref{pretable},  
\textbf{Column} $|\Delta|+|\Delta_c|$  is the number of  transitions of the SM-PDS  (changing and non  changing rules). 
\textbf{Column} SM-PDS gives  the cost it takes to apply our direct algorithm to compute the $pre^*$ for the given  SM-PDS.
\textbf{Column} PDS shows the cost it takes to get the equivalent PDS from the SM-PDS.
\textbf{Column} Symbolic PDS reports the cost it takes to get the equivalent Symbolic PDS from the SM-PDS.
\textbf{Column} Result1 reports the cost it takes to get the $pre^*$ analysis of Moped \cite{Schwoon:2007vs} for the  PDS we got.  
\textbf{Column} Total1 is the total cost it takes to translate the SM-PDS into a PDS and then apply the standard $pre^*$ algorithm of Moped (Total1=PDS+Result1).  
\textbf{Column} Result2 reports the cost it takes to get the $pre^*$ analysis of Moped for the symbolic PDS we got. 
\textbf{Column} Total2 is the total cost it takes to translate the SM-PDS into a symbolic PDS and then apply the standard $pre^*$ algorithm of Moped 
(Total2=Symbolic PDS+Result2). 
"error" in the table means failure of  Moped, because the size of the relations involved in the symbolic transitions  is  huge. 
Hence, we   mark  $-$ for the total execution time. 
You can see that our direct algorithm (\textbf{Column} SM-PDS) is much more efficient. 

Table \ref{posttable} shows the performance of our $post^*$ algorithm.  The meaning of the columns are exactly the same as for the $pre^*$ case, but using the $post^*$ algorithms instead. You can see from this table that applying our direct $post^*$ algorithm 
on the SM-PDS is much better than translating the SM-PDS to an equivalent PDS or symbolic PDS,  and then  applying the standard 
$post^*$ algorithms of Moped.  Going through PDSs or symbolic PDSs is less efficient and leads to memory out in several cases.

\subsection{Malware Detection}
Self-modifying code is widely used as an obfuscation technique for malware writers. 
Thus, we applied our tool for malware detection.
\begin{table}[htbp]
	\centering
	\begin{tabular}{|c|c|c|}
		\hline
		Example & SM-PDS &  PDS\\
		\hline
		Email-Worm.Win32.Klez.b& Y& N\\
		\hline
		Backdoor.Win32.Allaple.b&Y & N\\
		\hline
		Email-Worm.Win32.Avron.a &Y &N\\
		\hline
		Email-Worm.Win32.Anar.a &Y &N\\
		\hline
		Email-Worm.Win32.Anar.b &Y &N\\
		\hline
		Email-Worm.Win32.Bagle.a&Y &N\\
		\hline
		Email-Worm.Win32.Bagle.am&Y &N\\
		\hline
		Email-Worm.Win32.Bagle.ao &Y &N\\
		\hline
		Email-Worm.Win32.Bagle.ap  &Y &N\\
		\hline
		Email-Worm.Win32.Ardurk.d &Y &N\\
		\hline
		Email-Worm.Win32.Atak.k&Y &N\\
		\hline
		Email-Worm.Win32.Atak.g&Y &N\\
		\hline
		Email-Worm.Win32.Hanged &Y &N\\
		\hline
	\end{tabular}
	\caption{Malware Detection}
	\label{malwares}
\end{table}

We consider  self-modifying  versions of 13 well known malwares.
In these versions,  the malicious behaviors are unreachable if one does not take into account that the self-modifying piece of code will change the malware code: if the code does not change, the part that contains the malicious behavior cannot be reached; after executing the self-modifying code, the control point will jump to the part containing  the malicious behavior.

We model such malwares in two ways: (1)  first, we take into account the self-modifying piece of code and use SM-PDSs to represent these programs as discussed in Section \ref{prog-pds}, (2) second, we don't take into account that  this part of the  code is self-modifying and we treat it as all the other instructions of the program. In this case, we  model these programs by a standard PDS following the translation of 
\cite{SongT12}.

The results are reported in Table 3, 
\textbf{Column} Example reports the name of the  worm. \textbf{Column} SM-PDS shows the result obtained by  applying our method to check the reachability of  the entry point of the malicious block. \textbf{Column} PDS gives  the result if we apply the  traditional PDS translation of programs (without taking into account the semantics of self modifying code) method  to check the reachability of the entry point of the malicious block. $Y$ stands for yes (the program is malicious)  and $N$ stands for no (the program is benign). 
As it can be seen, our techniques that go through SM-PDS to model self modifying code is able to conclude that the entry point of the malicious block is 
reachable,  whereas the standard PDS translation from programs fails to reach this conclusion. 
\bibliographystyle{alpha}
\bibliography{paper.bib}

%



\end{document}